\documentstyle[12pt,aasms4]{article}

\slugcomment{Submitted to ApJ, Part I}

\lefthead{Gibson {\rm et~al. }}
\righthead{HST Key Project: The Distance to NGC~4725}

\begin{document}

\title{The $HST$ Key Project on the Extragalactic Distance Scale XVII. The 
Cepheid Distance to
NGC~4725\footnote{Based on observations with the 
NASA/ESA \it Hubble Space Telescope\rm, obtained at the Space Telescope 
Science Institute, which is operated by AURA, Inc., under NASA Contract No. 
NAS 5-26555.}}

\author{Brad K. Gibson\altaffilmark{2}, 
Shaun M.G. Hughes\altaffilmark{3},
Peter B. Stetson\altaffilmark{4}, 
Wendy L. Freedman\altaffilmark{5}, 
Robert C. Kennicutt, Jr.\altaffilmark{6},
Jeremy R. Mould\altaffilmark{7},
Fabio Bresolin\altaffilmark{8},
Laura Ferrarese\altaffilmark{9}, 
Holland C. Ford\altaffilmark{10}, 
John A. Graham\altaffilmark{11}, 
Mingsheng Han\altaffilmark{12},
Paul Harding\altaffilmark{6},
John G. Hoessel\altaffilmark{13}, 
John P. Huchra\altaffilmark{14}, 
Garth D. Illingworth\altaffilmark{15}, 
Daniel D. Kelson\altaffilmark{11},
Lucas M. Macri\altaffilmark{14},
Barry F. Madore\altaffilmark{16},
Randy L. Phelps\altaffilmark{17}, 
Charles F. Prosser\altaffilmark{18,19},
Abhijit Saha\altaffilmark{18}, 
Shoko Sakai\altaffilmark{18},
Kim M. Sebo\altaffilmark{7},
Nancy A. Silbermann\altaffilmark{16} and
Anne M. Turner\altaffilmark{6}}

\altaffiltext{2}{Center for Astrophysics and Space Astronomy, University of Colorado, Campus Box 389, Boulder, CO, USA  80309}
\altaffiltext{3}{Institute of Astronomy, Madingley Road., Cambridge, UK  CB3~0HA}
\altaffiltext{4}{Dominion Astrophysical Observatory, Herzberg Institute of Astrophysics, National Research Council, 5071 West Saanich Rd., Victoria, BC, Canada  V8X~4M6} 
\altaffiltext{5}{The Observatories, Carnegie Institution of Washington, Pasadena, CA, USA  91101}
\altaffiltext{6}{Steward Observatory, Univ. of Arizona, Tucson, AZ, USA  85721}
\altaffiltext{7}{Mount Stromlo \& Siding Spring Observatories, Australian National University, Weston Creek Post Office, Weston, ACT, Australia  2611}
\altaffiltext{8}{European Southern Observatory, D-85748 Garching b. M\"unchen, Germany}
\altaffiltext{9}{Palomar Observatory, California Institute of Technology, Pasadena, CA, USA  91125} 
\altaffiltext{10}{Dept. of Physics \& Astronomy, Bloomberg 501, Johns Hopkins Univ., 3400 N. Charles St., Baltimore, MD, USA  21218}
\altaffiltext{11}{Dept. of Terrestrial Magnetism, Carnegie Institution of Washington, 5241 Broad Branch Rd. N.W., Washington, D.C., USA  20015}
\altaffiltext{12}{Avanti Corporation, 46871 Bayside Parkway, Fremont, CA, USA 94538}
\altaffiltext{13}{Univ. of Wisconsin, Madison, WI, USA, 53706}
\altaffiltext{14}{Harvard College, Center for Astrophysics, 60 Garden St., Cambridge, MA, USA  02138}
\altaffiltext{15}{Lick Observatory, Univ. of California, Santa Cruz, CA, USA 95064}
\altaffiltext{16}{Infrared Processing and Analysis Center, Jet Propulsion Laboratory, California Institute of Technology, Pasadena, CA, USA  91125}
\altaffiltext{17}{Wright Laboratory of Physics, Oberlin College, 110 North Professor St., Oberlin, OH, USA  44074-1088}
\altaffiltext{18}{National Optical Astronomy Observatories, P.O. Box 26732, Tucson, AZ, USA  85726}
\altaffiltext{19}{Deceased}

\def\spose#1{\hbox to 0pt{#1\hss}}
\def\simlt{\mathrel{\spose{\lower 3pt\hbox{$\mathchar"218$}}
     \raise 2.0pt\hbox{$\mathchar"13C$}}}
\def\simgt{\mathrel{\spose{\lower 3pt\hbox{$\mathchar"218$}}
     \raise 2.0pt\hbox{$\mathchar"13E$}}}
\def\eg{{\rm e.g. }}
\def\ie{{\rm i.e. }}
\def\etal{{\rm et~al. }}

\begin{abstract}
The distance to NGC~4725 has been derived from Cepheid variables, as part of 
the \it Hubble Space Telescope
Key Project on the Extragalactic Distance Scale\rm.  Thirteen 
F555W (V) and four F814W (I) epochs of cosmic-ray-split Wide Field 
and Planetary Camera 2 observations were obtained.  Twenty Cepheids were
discovered, with periods ranging from 12 to 49 days.  
Adopting a Large Magellanic Cloud distance modulus and extinction of
$18.50\pm 0.10$ mag and E(V$-$I)=0.13 mag, respectively, a true 
reddening-corrected distance modulus (based on an analysis employing the
ALLFRAME software package)
of $30.50\pm 0.16 \,({\rm random}) \pm 0.17 \,({\rm systematic})$ 
mag was determined for NGC~4725.  The corresponding of 
distance of $12.6\pm 1.0 \,({\rm random}) \pm 1.0
\,({\rm systematic})$ Mpc is in excellent agreement with that found with
an independent analysis based upon the DoPHOT photometry package.  With a
foreground reddening of only E(V$-$I)=0.02, the inferred \it 
intrinsic \rm reddening of this field in NGC~4725, E(V$-$I)=0.19, makes 
it one of
the most highly-reddened, encountered by the \it HST Key Project\rm, to date.
\end{abstract}

\keywords{Cepheids --- distance scale --- galaxies: distances and redshifts ---
galaxies: individual (NGC~4725)}

\section{Introduction}
\label{introduction}

The \it Hubble Space Telescope (HST) Key Project on the Extragalactic Distance
Scale \rm has as its primary goal the determination of the Hubble constant to
an accuracy $\simlt 10$\% (Kennicutt, Freedman \& Mould 1995).  Cepheid
distances to 18 spirals, within $\sim 20$ Mpc, are being obtained and will
be used to calibrate a variety of secondary distance indicators, including the
Tully-Fisher relation (TF), surface brightness fluctuations (SBF), 
planetary nebula luminosity function (PNLF), globular cluster luminosity 
function (GCLF), and Type Ia supernovae.

NGC~4725 is an Sb/SB(r)II barred spiral (Sandage 1996), with an uncorrected HI
21cm linewidth of $\sim 411$ km/s (Wevers \etal 1984), and an isophotal
inclination of $\sim 46^\circ$ (de Vaucouleurs \etal 1991 - although, see
Section \ref{irtf}). 
Its position ($\alpha=12^{\rm h}50^{\rm m}27^{\rm
s},\delta=+25^\circ 30^\prime 06^{\prime\prime}$, J2000) and
Galactocentric radial velocity $v=1207$ km/s (de Vaucouleurs \etal 1991)
led to its assignment to the Coma-Sculptor Cloud.
NGC~4725 and 4747 are relatively
isolated dynamically from the remainder of the Cloud (\eg Zaritsky \etal 1997),
and comprise what has
come to be known as the Coma II Group of galaxies (\eg Table II of Tully
1988).  NGC~4725 is one of the \it HST Key Project \rm
primary calibrators for the infrared
Tully-Fisher (IRTF) relationship.  Because of the (assumed) association of
the Coma II Group with that of the neighboring (larger)
Coma I Group\footnote{After Tully (1988),
the Coma I Group is comprised of 25 members, including the notable elliptical
NGC~4494 and edge-on sprial NGC~4565.}
(and, to some degree, the
Coma-Sculptor Cloud as a whole), it was hoped that NGC~4725 would indirectly
provide calibration for the SBF, PNLF, and GCLF secondary candles, a 
point to which we return in Section \ref{distcompare}.  
NGC~4725 was the host galaxy for supernova SN1940B, 
a typical example of the ``regular'' class of ``plateau'' Type II events
(Patat \etal 1994), but data do not exist which would allow application of
the expanding photosphere method secondary distance indicator.

In Section \ref{photometry} we present our multi-epoch Wide Field and
Planetary Camera 2 (WFPC2) \it HST \rm observations and review the 
two independent approaches taken to the photometry and calibration of
the instrumental magnitudes -- the methodology employed follows that of previous
papers in this series (\eg Stetson \etal 1998, and
references therein).  The identification of Cepheids and their derived
properties, again employing two independent algorithms, are discussed in Section
\ref{cepheids}.  The derived distance to NGC~4725 is presented in Section
\ref{distance}, and the result contrasted with previous distance determinations
for NGC~4725 and the Coma I/II galaxy groups, in Section \ref{distcompare}.
A summary is provided in Section \ref{summary}.

\section{Observations and Photometry}
\label{photometry}

\it HST \rm WFPC2 observations of NGC~4725 were carried out over a two month
period (1995 April 12 - 1995 June 14), 
with a single V epoch revisit on 1996 April 29.  The revisit epoch,
approximately one year after the conclusion of the main observing window, was
used to constrain the periods of the longer-period Cepheids.  In total,
thirteen epochs of F555W (V), four epochs of F814W (I), and two
epochs of F439W (B) were covered.  Each epoch consisted of a pair of
cosmic-ray-split exposures, each of duration 1000-1500 s.  
Because of the sparse phase
coverage of the F439W observations and, more importantly, their insufficient 
signal-to-noise for detecting the majority of the Cepheid candidates, these 
were not included in the analysis which
follows.  The observing strategy, optimized to uncover Cepheids with periods
$\sim 10-60$ days, follows that outlined in Freedman \etal (1994).  The
individual epochs, \it HST \rm archive filenames, time at which a given epoch's
observations began, and the exposure times and filters employed, are all listed
in Table \ref{tbl:observations}.

\placetable{tbl:observations}

A $10^\prime\times 10^\prime$ ground-based
image, obtained with the 2.5m Isaac Newton Telescope, is
shown in Figure \ref{fig:footprint};  the WFPC2 footprint has been
superimposed.  WFPC2 incorporates four $800\times 800$ CCDs; the 
Planetary Camera (PC) has a $37\times 37$ arcsecond field of view, and is
referred to as Chip 1, while the three Wide Field Camera (WFC) chips have 
$80\times 80$
arcsecond fields of view each, and are referred to as Chips 2, 3, and 4, 
respectively, moving counter-clockwise from the PC in Figure
\ref{fig:footprint}.

\placefigure{fig:footprint}

As in previous papers in this series, dual independent analyses were undertaken
using ALLFRAME (Stetson 1994) and DoPHOT (Saha \etal 1996, and references
therein).  As detailed descriptions of the reduction process can be found in
Stetson \etal (1998), we only provide a
brief summary of the key steps, in what follows.

\subsection{ALLFRAME}
\label{allframe}

The input star list to ALLFRAME was generated by median averaging the 26 F555W
and 8 F814W cosmic-ray-split
images of Table \ref{tbl:observations} to produce cosmic-ray-free
frames for each chip/bandpass combination.  
Iterative application of DAOPHOT and ALLSTAR led to the
final master star list, which was input to ALLFRAME, and used to extract
profile-fitting stellar photometry from the 34 individual frames.  The adopted
point spread functions (PSFs) were derived from public domain
\it HST \rm WFPC2 observations of the globular clusters Pal~4 and NGC~2419.

Aperture photometry was performed on the 45 isolated bright stars listed in
Table \ref{tbl:refstars}. The program DAOGROW was then
employed to generate growth curves out to $0^{\prime\prime}$\llap{.}5, 
allowing an aperture correction to be derived for each chip and filter, to
ensure a match to the Holtzmann \etal (1995) photometric system.
The photometric zero points, aperture corrections, and
long-exposure zero-point correction were then used to finally convert from
instrumental magnitudes to the standard system, following the procedure
outlined in Stetson \etal (1998).

\placetable{tbl:refstars}

\subsection{DoPHOT}
\label{dophot}

The DoPHOT philosophy concerning treatment of cosmic rays
differs from that of ALLFRAME, in that each cosmic-ray-split pair was first
combined using a sigma detection algorithm which takes into account the
problems of undersampling (Saha \etal 1996).  The final calibration of DoPHOT
magnitudes follows that detailed in Section 2.2 of Stetson \etal
(1998). Instrumental magnitudes were corrected to a $0^{\prime\prime}$\llap{.}5
aperture magnitude using aperture corrections and zero points appropriate for
long exposures, and converted to the standard system (Holtzmann \etal 1995).
Calibrated DoPHOT photometry (and the associated error),
for the 45 NGC~4725 reference stars, is listed in Table \ref{tbl:refstars}.

\subsection{Comparison Between ALLFRAME and DoPHOT Photometry}
\label{comparison}

A chip-by-chip comparison of ALLFRAME and DoPHOT photometry (both 
V- and I-bands) for the 45 reference stars of Table \ref{tbl:refstars} is 
provided in Table \ref{tbl:all_do_comp}.  The agreement is very good for Chips
2-4 (\ie the WFC fields), with a mean difference of $-0.01\pm 0.07$ mag in V,
and $-0.03\pm 0.07$ mag in I, being determined (in the sense of
ALLFRAME-DoPHOT).
The largest single chip$+$filter discrepancy found is
$-0.07\pm 0.07$ mag in I for Chip 4, which considering the 15 reference stars
employed, is discrepant at the $\sim 4\sigma$ level.  Such residual
offsets have been observed in all \it HST Key Project \rm galaxies to date; 
artificial star tests are currently underway, in order to
ascertain and quantify the source of these discrepancies (Ferrarese \etal 1999).
Due to the absence of Cepheid candidates and bright reference stars in
Chip 1 (\ie the PC field), the comparison of Table \ref{tbl:all_do_comp} 
is restricted to Chips 2-4.

\placetable{tbl:all_do_comp}

The comparison between ALLFRAME and DoPHOT mean magnitudes, for each of the 20
Cepheid candidates (detailed in
Section \ref{cepheids}), is
likewise presented in Table \ref{tbl:all_do_comp}.  The mean differences of
$+0.057\pm 0.020$ mag in V, and $+0.016\pm 0.017$ mag in I, are manifest
in the slight offsets
between the ALLFRAME and DoPHOT period-luminosity (PL) fits
noted in Section \ref{distance}.

\section{Cepheid Identification}
\label{cepheids}

In a similar vein to the philosophy of performing dual independent photometric
reductions with ALLFRAME and DoPHOT, independent 
Cepheid identification techniques were employed by each reduction team.
Candidate Cepheids were extracted from the ALLFRAME dataset using TRIAL,
Stetson's (1996) template light curve fitting algorithm, whereas a variant of
Stellingwerf's (1978) phase dispersion minimization routine (Hughes 1989,
and referred to as PDM henceforth) was adopted for the DoPHOT dataset.

Twenty high quality candidates were uncovered, the assigned
identification numbers and coordinates 
(both (X,Y) on the respective WFC chip and (RA,DEC)) for which are listed
in Table \ref{tbl:cepheids}.  The spatial distribution of the Cepheids in each
chip is shown in Figure \ref{fig:fov}, with 
detailed ($4^{\prime\prime}\times 4^{\prime\prime}$ windows centered upon each
Cepheid) finding charts available in Figure \ref{fig:charts}.

\placetable{tbl:cepheids}

\placefigure{fig:fov}

\placefigure{fig:charts}

The corresponding period and mean magnitude for each of the 20 Cepheids in
question, as reported by TRIAL (for ALLFRAME data) and PDM (for DoPHOT data),
is reproduced (along with their accompanying errors) in Table
\ref{tbl:cepheids2};  ALLFRAME
light curves for each, phased to their respective period,
are presented in Figure \ref{fig:curves} - V- and I-band photometry
represented by solid dots and open squares, respectively.  The tabulated
epoch-by-epoch ALLFRAME photometry (and associated errors), for each of the
Cepheids, is given in Table \ref{tbl:magnitudes}.  We have chosen to present 
all the photometry in Figure \ref{fig:curves}, including those epochs
obviously affected by cosmic ray hits for Cepheids C04, C06, C08, C11,
C12, and C17.  
It is important to stress though that the periods and mean magnitudes
assigned by TRIAL and PDM have not been affected by these outliers, as
clearly outlined by Stetson (1996) and Stetson \etal (1998).

\placetable{tbl:cepheids2}

\placefigure{fig:curves}

\placetable{tbl:magnitudes}

The 20 Cepheids listed in Table \ref{tbl:cepheids2} have been
identified in the deep V versus V$-$I color-magnitude diagram (CMD) of Figure
\ref{fig:cmd}; all clearly lie in the instability strip.

\placefigure{fig:cmd}

\section{The Distance to NGC~4725}
\label{distance}

As described previously by Ferrarese \etal (1996), the apparent V- and I-band
distance moduli (\ie $\mu_{\rm V}$ and $\mu_{\rm I}$)
to NGC~4725 are derived relative to that of the LMC, adopting
Madore \& Freedman's (1991) LMC PL relations, scaled to a true modulus of
$\mu_\circ=18.50\pm 0.10$ mag and reddening E(V$-$I)=0.13.
In fitting to the NGC~4725 Cepheid data (Table \ref{tbl:cepheids2}), the slopes
of the PL relations were fixed to those of Madore \& Freedman's LMC PL
relations.  

The ALLFRAME/DoPHOT V- and I-band PL relations for NGC~4725 are shown in
Figures \ref{fig:pl_all} and \ref{fig:pl_do}.  
The 20 Cepheids used in the final regression are
denoted with solid circles, and listed in Table \ref{tbl:cepheids2}.  The solid
lines shown are the best fit regression, imposing the LMC PL slopes, 
while the dotted lines represent 2$\sigma$ deviations from the mean of the LMC
relations (\ie 0.54 mag in V, and 0.36 mag in I - Madore \& Freedman 1991).
The resulting apparent ALLFRAME 
distance moduli are $\mu_{\rm V}=31.00\pm 0.06$ mag and
$\mu_{\rm I}=30.80\pm 0.06$ 
mag, with DoPHOT values of $\mu_{\rm V}=30.95\pm 0.07$ mag
and $\mu_{\rm I}=30.79\pm 0.06$ mag.  
The 0.05 and 0.01 mag offsets in the apparent ALLFRAME and DoPHOT V-and 
I-band distance moduli, respectively, simply reflect the 0.057 and 0.016 mag 
Cepheid mean magnitude offsets presented in Section \ref{comparison}.
The derived reddenings are
E(V$-$I)=0.21$\pm$0.02 (ALLFRAME) and E(V$-$I)=0.16$\pm$0.03 (DoPHOT).
The DIRBE/IRAS dust maps of Schlegel, Finkbeiner \& Davis (1998)
show a foreground reddening component
of only E(V$-$I)=0.02 along this sight line; of the 19 galaxies examined by the
\it HST Key Project \rm to date, 
NGC~4725 possesses the greatest internal extinction.\footnote{Indeed, 
both NGC~4725
and 3621 are $2\rightarrow 3\sigma$ outliers, in relation to the sample
used by Kennicutt \etal (1998) 
to derive the relationship between Cepheid color excess and
metallicity (\ie $\delta{\rm E(V-I)}/\delta[{\rm O/H}]=0.12\pm 0.08$ mag/dex),
in the sense of having higher excesses than expected for their metallicity.}

\placefigure{fig:pl_all}

\placefigure{fig:pl_do}

An independent estimate of the extinction internal to NGC~4725 can be
obtained using the HI maps of Wevers \etal (1984), from which the 
HI surface density of the WFPC2 field is $\sim 6\pm 2 \times 10^{20}$
atoms cm$^{-2}$.  This range in HI column density should have associated 
with it enough dust to produce color excesses in the range E(V$-$I)=0.03 
to 0.14, assuming a similar gas/dust ratio for NGC~4725 as for the low 
extinction regions of the Milky Way (\ie equation 7 of
Burstein \& Heiles 1978).
Our mean extinction for the Cepheids, although larger than this, is in 
reasonable agreement given the large uncertainties associated with gas-to-dust
ratios, which in the Milky Way is reflected in the Heiles (1976)
``R-parameter'' (a useful measure of the total Galactic
gas to dust ratio - see also equation 4 of Burstein \& Heiles) ranging
from -6 to +6, corresponding to a factor of 4 range in dE(B$-$V)/dN$_{\rm HI}$
(Burstein \& Heiles 1978).

Adopting a ratio of total to selective absorption of A$_{\rm V}=2.45\,{\rm
E(V-I)}$, consistent with the Cardelli \etal (1989) extinction law, we derive 
true ALLFRAME and DoPHOT distance moduli of $\mu_\circ=30.50\pm 0.06$ mag and
$\mu_\circ=30.55\pm 0.07$ mag, respectively, corresponding to $d=12.6\pm 0.4$
Mpc and $d=12.9\pm 0.4$ Mpc.\footnote{The PL fitting methodology adopted in the
current study is identical to that adopted throughout the $HST$ Key Project
series - \ie the absolute distance modulus $\mu_\circ$, associated reddening
E(V$-$I), and their associated uncertainties, are derived directly
from the full sample of 20 individually de-reddened Cepheids, and not 
indirectly through a comparison of the apparent distance moduli ($\mu_{\rm V}$
and $\mu_{\rm I}$).}
Restricting ourselves to the 15 Cepheids
with P$>$20d, as a test for incompleteness bias in the sample, the inferred
distance moduli increase by 0.07 mag - an $\sim 1\sigma$ effect - allowing
us to conclude that our results have not been severely compromised by such
a bias.

The errors listed above reflect internal errors alone, arising from
scatter in the NGC~4725 PL relations.  A more complete assessment of the
associated uncertainty, incorporating other potential random and systematic 
errors, is presented in Table \ref{tbl:error}. Uncertainties due to
metallicity, LMC distance modulus, and photometric calibration all contribute
to the NGC~4725 distance modulus error budget.  

As in previous papers in this series (\eg Hughes \etal 1998), the systematic
uncertainty introduced by the adopted Cepheid PL calibration - $\pm 0.12$ mag 
(S1 in Table \ref{tbl:error}) - is dominated by the error in the LMC true
modulus ($\pm 0.10$ mag, from Madore \& Freedman 1991 and Westerlund 1996).  

The remaining systematic uncertainty in Table \ref{tbl:error} which should be
noted here is that due to a possible metallicity dependence of the
Cepheid PL relation at V and I.  Kennicutt \etal (1998), based upon two fields
in M101, find a metallicity dependence of the form
d$\mu_\circ$/d[O/H] = -0.24$\pm$0.16 mag/dex.  If it can be shown that the NGC
4725 Cepheids differ substantially in metal abundance from those of the LMC
Cepheids which calibrate the PL relation, a significant systematic error could
be introduced into the derived distance.

Based upon 8 HII regions in NGC~4725, Zaritsky \etal (1994) determined
a mean oxygen abundance of $12+\log({\rm O/H})=9.26\pm 0.57$, at a
galactocentric distance $r=3$ kpc, with a corresponding abundance gradient of
$-0.022\pm 0.063$ dex/kpc.  With the WFPC2 fields at a
galactocentric distance
of $\sim 13\pm 2$ kpc (recall the $10^\prime\times 10^\prime$ scale of 
Figure \ref{fig:footprint}), we therefore
estimate a mean oxygen abundance for the fields of $12+\log({\rm
O/H})\approx 9.0\pm 0.3$.
In contrast, the mean calibrating
LMC HII region abundance used by the \it HST Key Project \rm
is $12+\log({\rm O/H})=8.5$ (Kennicutt \etal 1998).  Recalling the
aforementioned Kennicutt \etal Cepheid metallicity dependence, this
possible
factor of three greater Cepheid metallicity in NGC~4725, in comparison with the
LMC Cepheids, could cause the Cepheid
distance modulus to NGC~4725 to be underestimated by $\sim 0.12\pm 0.21$ mag.
In keeping with earlier papers in this series, and noting the large uncertainty
attached to the metallicity extrapolation for our Cepheid field, this potential 
correction to the distance modulus of $+0.12\pm 0.21$ mag is not currently 
applied, but simply added to the appropriate systematic error budget in 
Table \ref{tbl:error}.  We will revisit the issue of the effects of metallicity
on distances to \it HST Key Project \rm
galaxies in a consistent manner when all of 
the measurements of galaxy distances in our sample have been completed.

In light of the complete list of random and systematic errors shown in Table
\ref{tbl:error}, our final quoted Cepheid-based true distance moduli
to NGC
4725 are $\mu_\circ=30.50\pm 0.16\,({\rm random})\pm 0.17\,
({\rm systematic})$ mag
(ALLFRAME) and $\mu_\circ=30.55\pm 0.16\,({\rm random})\pm 0.17\,({\rm
systematic})$ mag (DoPHOT), with reddenings of E(V$-$I)=0.21$\pm$0.02 (internal)
and E(V$-$I)=0.16$\pm$0.03 (internal), respectively.  
The corresponding distances are $12.6\pm 1.0\,({\rm
random})\pm 1.0\,({\rm systematic})$ Mpc (ALLFRAME)
and $12.9\pm 1.0\,({\rm random})\pm 1.0\,({\rm systematic})$ Mpc (DoPHOT).

\section{Previous Distance Determinations for NGC~4725 and the Coma Cloud}
\label{distcompare}

Previous distance determinations for NGC~4725 have been based upon either
measurements of NGC~4725 itself, or indirectly through an assumed association
with the Coma I or II Groups, within the Coma-Sculptor Cloud.
Table \ref{tbl:distances} provides a summary of both families of distance
determinations.  The method employed is noted in column 1, the distance and
quoted error (both in Mpc) in column 2, and the appropriate reference in column
3.  Again, Tully's (1988) Coma Cloud inventory has been adopted
in Table \ref{tbl:distances}; while convenient, this should \it not \rm be
construed as unequivocal support for any assumed physical association 
between these tabulated galaxies.

\placetable{tbl:distances}

Early quoted values for NGC~4725 proper
include Bottinelli et~al.'s (1985) B-band Tully-Fisher (TF)-derived value of
$d=9.9\pm 1.0$ Mpc and 
Tully's (1988) $d=12.4$ Mpc, derived from assuming H$_\circ=75$ km/s/Mpc, along
with a simple Virgocentric flow model.  Subsequent to this, and adopting an
H-band TF relationship zero-point tied to M31, M33, and NGC~2403, 
Tully \etal (1992) found $d=16.1$ Mpc.\footnote{Note, though, that
the predicted distance output
from their mass model of the same paper (Tully \etal 1992) was 20 Mpc,
symptomatic of the well-documented ``triple-value ambiguity'', discussed
therein.}  Tully (1997)
has since revised the TF distance to NGC~4725, by taking into account not only
the H band, but also B-, R-, and I-bands, the average of which yields
$12.6\pm 2.1$ Mpc, in 
agreement with our Cepheid distance of $12.6\pm 1.0\,({\rm random})\pm 1.0\,(
{\rm systematic})$ Mpc.

NGC~4725 is one of only 6 spirals for which a \it direct \rm
comparison between SBF- and either maser- or 
Cepheid-derived distances can be made
(the others being NGC~224, 3031, 3368, 4258, and 7331).
Tonry's (1998) interim SBF distance to the bulge of NGC~4725 of
$13.1\pm 2.2$ Mpc, likewise agrees with our newly-derived
Cepheid distance.\footnote{Tonry's
(1998) interim distance to the bulge of NGC~4725 (\ie $13.1\pm 2.2$ Mpc)
supersedes the earlier \it indirect \rm SBF measurement to the Coma II
Group of galaxies of $15.9\pm 0.6$ Mpc 
(Tonry \etal 1997).  The latter was based upon an assumed association with
NGC~4494 and 4565, an assumption no longer necessary with the now available
\it direct \rm SBF distance determination.  As recognized by
Tonry et~al., the disagreement 
between SBF and PNLF/GCLF distances to NGC~4494 and 4565 (Fleming \etal 1995; 
Forbes 1996; Jacoby \etal 1996) still remains unresolved, and is reflected in
the relevant entries to Table \ref{tbl:distances}.}  The SBF Survey is
currently undergoing final recalibration (Tonry 1998), at which time the
definitive comparison can be made.

The range of direct and indirect distance determinations to galaxies
generally associated with the classical Coma Groups I and II
(ie 10-19 Mpc), as reflected
by the compilation of Table \ref{tbl:distances}, reinforces the conclusions
of Turner \etal (1998, Section 7) 
regarding the limited use of these two galaxy groups
for calibrating secondary distance indicators.

\subsection{Infrared Tully-Fisher Relationship}
\label{irtf}

The interim H-band Tully-Fisher relation, derived 
from the 19 calibrators currently available,
is shown in Figure \ref{fig:irtf}, and follows that given by
Mould \etal (1997).  The Aaronson et~al. (1982)
H-band photometric index H$_{-0.5}$, coupled with the 21cm linewidths
tabulated by Tormen \& Burstein (1995), were employed.  For comparison,
Freedman's (1990) earlier IRTF calibration, based upon only 5 local 
calibrators, is shown.
The 0.47 magnitude offset between the two calibrations
can be traced to the sample of five galaxies available to Freedman in 1990;
\it each \rm 
of these five (NGC~224, 300, 598, 2403, and 3031) are $\sim 0.2\rightarrow
0.5$ mag fainter in H$_{-0.5}$ than the mean global values found for
expected for their respective HI
linewidths.  This only becomes apparent when the entire
sample of 19 IRTF calibrators is considered.  

\placefigure{fig:irtf}

The subject of this paper, NGC~4725, identified
in Figure \ref{fig:irtf} by the open circle,
possesses an H-band luminosity a factor of two
lower than expected for its linewidth (similar to NGC~224$\equiv$M31). 
While we do not wish to belabor or overinterpret this mild divergence 
($< 1\sigma$) from the mean Tully-Fisher relation, there are several points
which should be made.  It is apparent that NGC~4725 and its neighbor
NGC~4747 (at a projected distance of $\sim 88$ kpc) have undergone a past
encounter.  The striking $\sim 50$ kpc-long HI plumes extending from the center
of NGC~4747, including the one pointed directly at NGC~4725, clearly support
this picture (Wevers \etal 1984).  Given that NGC~4725 is twenty times as
massive as NGC~4747, it is not surprising to find that while the latter is
severely distorted, the former is far more stable against tidal interactions
and only shows a minor elongation and possible warping of the outer
south-eastern spiral arm.  Still, a consequence of this distorted spiral arm is
that the outer isophotes are less elongated than the inner ones, which may lead
to an underestimate of the inclination should it be based solely on the outer
isophotes.  This may be the 
source of the mild discrepancy between the photometric
inclination of 46$^\circ$ (de Vaucouleurs \etal 1991) and the outer disk HI
kinematic inclination of 53$^\circ$ (Wevers \etal 1984), although it should
be stressed that the values are consistent within the quoted errors ($\pm
4^\circ$).  We note in passing that increasing the assumed inclination from
$46^\circ$ to $53^\circ$ will have the effect of shifting the
$\log(\Delta{\rm V})$ for NGC~4725 in
Figure \ref{fig:irtf} from 2.76 to 2.71, eliminating its $\sim 1\sigma$ outlier
status from the mean IRTF relation.  Such issues will be addressed fully 
in the \it HST Key Project\rm's 
Tully-Fisher calibration paper (Sakai \etal 1998);
for the time being though, we retain complete
self-consistency with the compiled H-band magnitudes and 21cm linewidths in
Tormen \& Burstein (1995).

\section{Summary}
\label{summary}

\it HST \rm WFPC2 imaging of the Coma II group galaxy NGC~4725 has led to the
discovery of twenty Cepheids with periods ranging from 12 to 49
days.  Based upon the resultant V- and I-band period-luminosity relations, we
obtained true distance moduli of $30.50\pm 0.16 \,({\rm random}) \pm 0.17 
\,({\rm
systematic})$ and $30.55\pm 0.16 \,({\rm random}) \pm 0.17 
\,({\rm systematic})$ mags, and
reddenings of E(V$-$I)=$0.21\pm 0.02$ (internal) and $0.16\pm 0.03$ (internal)
mags, for
the ALLFRAME- and DoPHOT-reduced datasets, respectively.  The corresponding 
distances are then $12.6\pm 1.0 \,({\rm random}) \pm 1.0 \,({\rm systematic})$
and $12.9\pm 1.0 \,({\rm random}) \pm 1.0 \,({\rm systematic})$ 
Mpc, in excellent agreement with
the most recent Tully-Fisher (Tully 1997) and SBF (Tonry 1998)
distances to NGC~4725.  While useful as a calibrator for these two secondary
distance indicators, we echo the conclusions of Turner \etal (1998) in that 
our Cepheid-derived distance is of limited use for calibrating secondary 
indicators pertaining to the Coma-Sculptor Cloud proper.

\acknowledgments

The work presented in this paper is based on observations with the NASA/ESA
Hubble Space Telescope, obtained by the Space Telescope Science Institute,
which is operated by AURA, Inc. under NASA contract No. 5-26555.
The continued assistance of the NASA and STScI support staff, and in particular
our program coordinator, Doug Van Orsow, is gratefully acknowledged.
Support for this work was provided by NASA through grant GO-2227-87A from
STScI. SMGH and PSB are grateful to NATO for travel support via a Collaborative
Research Grant (960178).  We wish to thank John Tonry, Brent Tully, 
Bill Harris, and Robin Ciardullo, for many enlightening correspondences.

\clearpage

\figcaption[f1.eps]{
A $10^\prime\times 10^\prime$
ground-based image of NGC~4725, taken at the 2.5m Isaac Newton Telescope.
North is to the top and east to the left.
The WFPC2 footprint is superimposed, where C1 represents the Planetary Camera
chip, and C2, C3, and C4 the Wide Field Camera chips.  
{\bf See accompanying jpg file: 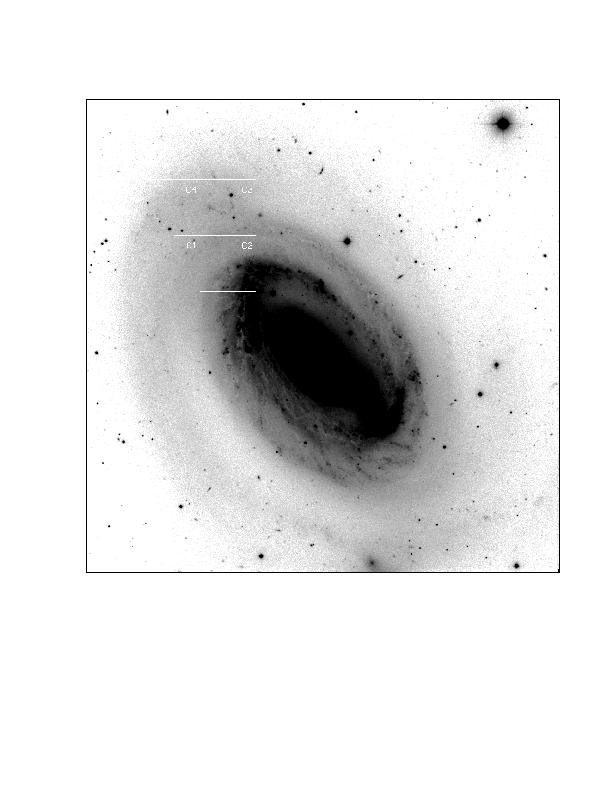}
\label{fig:footprint}}

\figcaption[f2a.eps]{
(a) The $80^{\prime\prime}\times 80^{\prime\prime}$ field of view of the WFC
Chip 2 in NGC~4725.  North is toward the left, east to the bottom.  Locations
of the Cepheid candidates are marked, with detailed finding charts available
for each in Figure \ref{fig:charts}.
(b) The $80^{\prime\prime}\times 80^{\prime\prime}$ field of view of the WFC
Chip 3 in NGC~4725.  North is toward the top, east to the left.  Locations
of the Cepheid candidates are marked, with detailed finding charts available
for each in Figure \ref{fig:charts}.
(c) The $80^{\prime\prime}\times 80^{\prime\prime}$ field of view of the WFC
Chip 4 in NGC~4725.  North is toward the right, east to the top.  Locations
of the Cepheid candidates are marked, with detailed finding charts available
for each in Figure \ref{fig:charts}.
{\bf See accompanying jpg files: 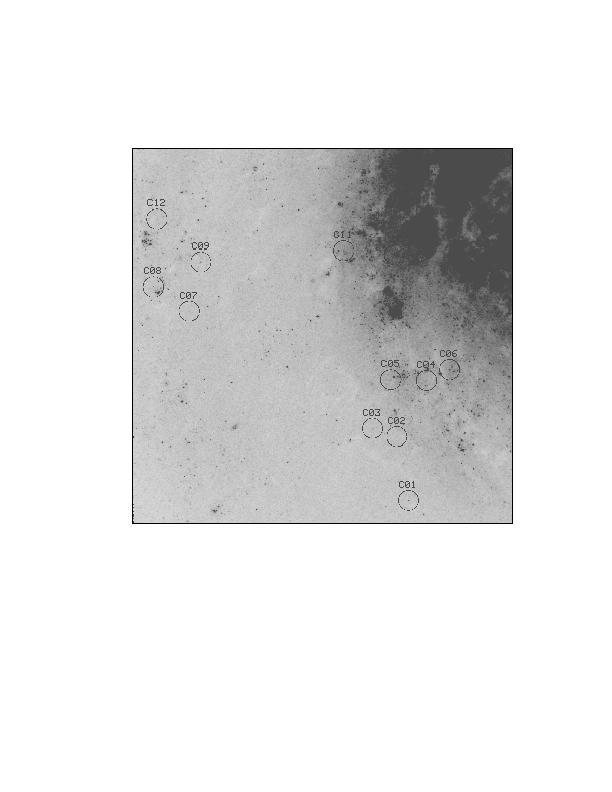 - 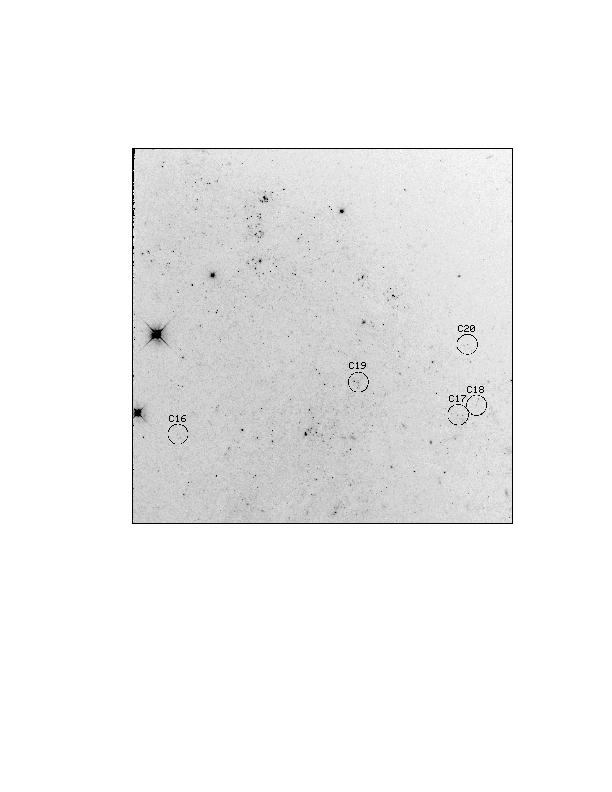}
\label{fig:fov}}

\figcaption[f3.eps]{
Finder charts for each of the Cepheid candidates for NGC~4725.  Each image is
$41\times 41$ pixels (\ie $4^{\prime\prime}\times 4^{\prime\prime}$), with an
orientation matching that of Figure \ref{fig:fov}.  In each case, the Cepheid
is situated at the exact center of the panel.
{\bf See accompanying jpg file: 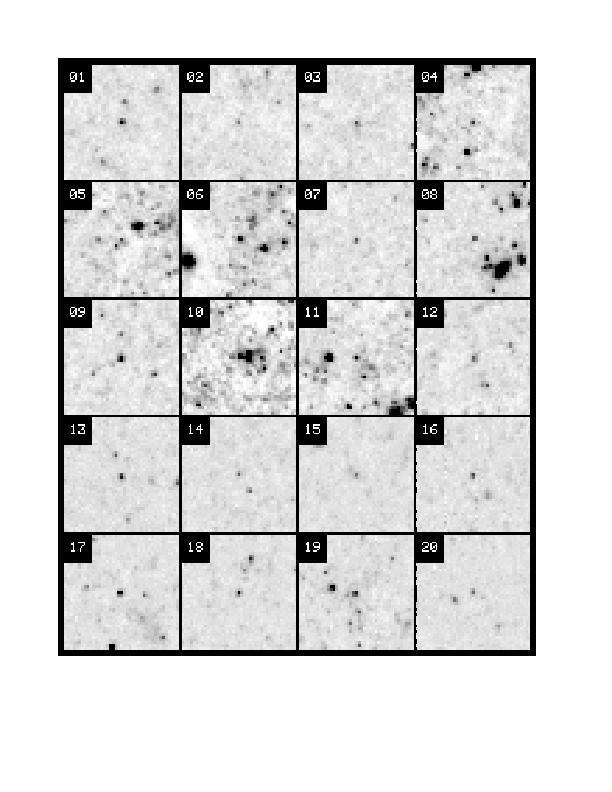}
\label{fig:charts}}

\figcaption[f4a.eps]{
Calibrated
ALLFRAME V- (filled circles) and I-band (open squares) phased lightcurves (two
cycles), for the Cepheids listed in Table \ref{tbl:cepheids}.  All data from 
Table \ref{tbl:magnitudes} are shown, including obvious cosmic-ray-affected
epochs in the lightcurves of Cepheids C04, C06, C08, C11, C12, and C17.  Said
cosmic-ray hits do not impact upon the period determination (Stetson 1996), but
are shown for completeness.
{\bf See accompanying jpg files: 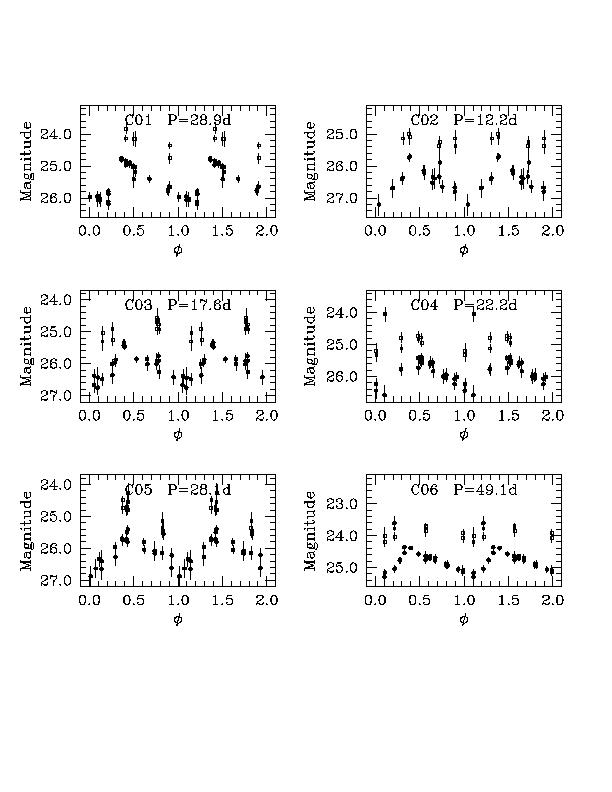 - 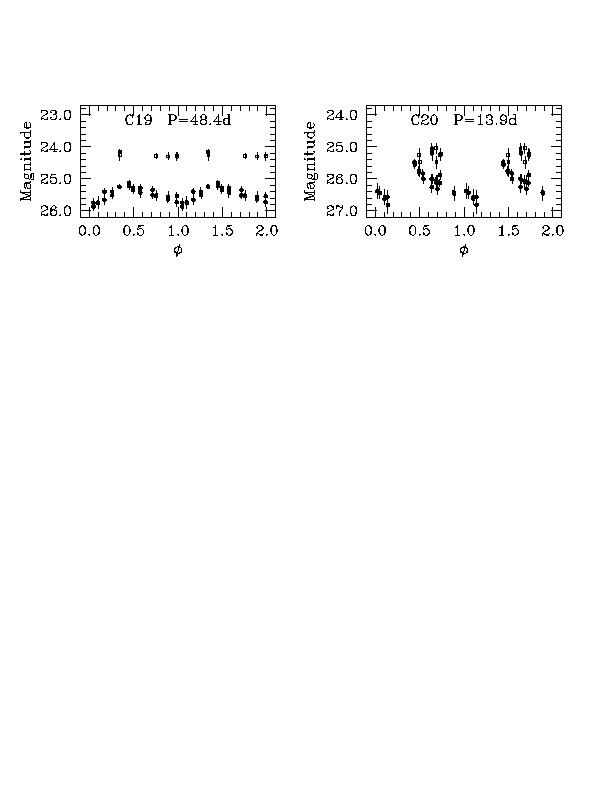}
\label{fig:curves}}

\figcaption[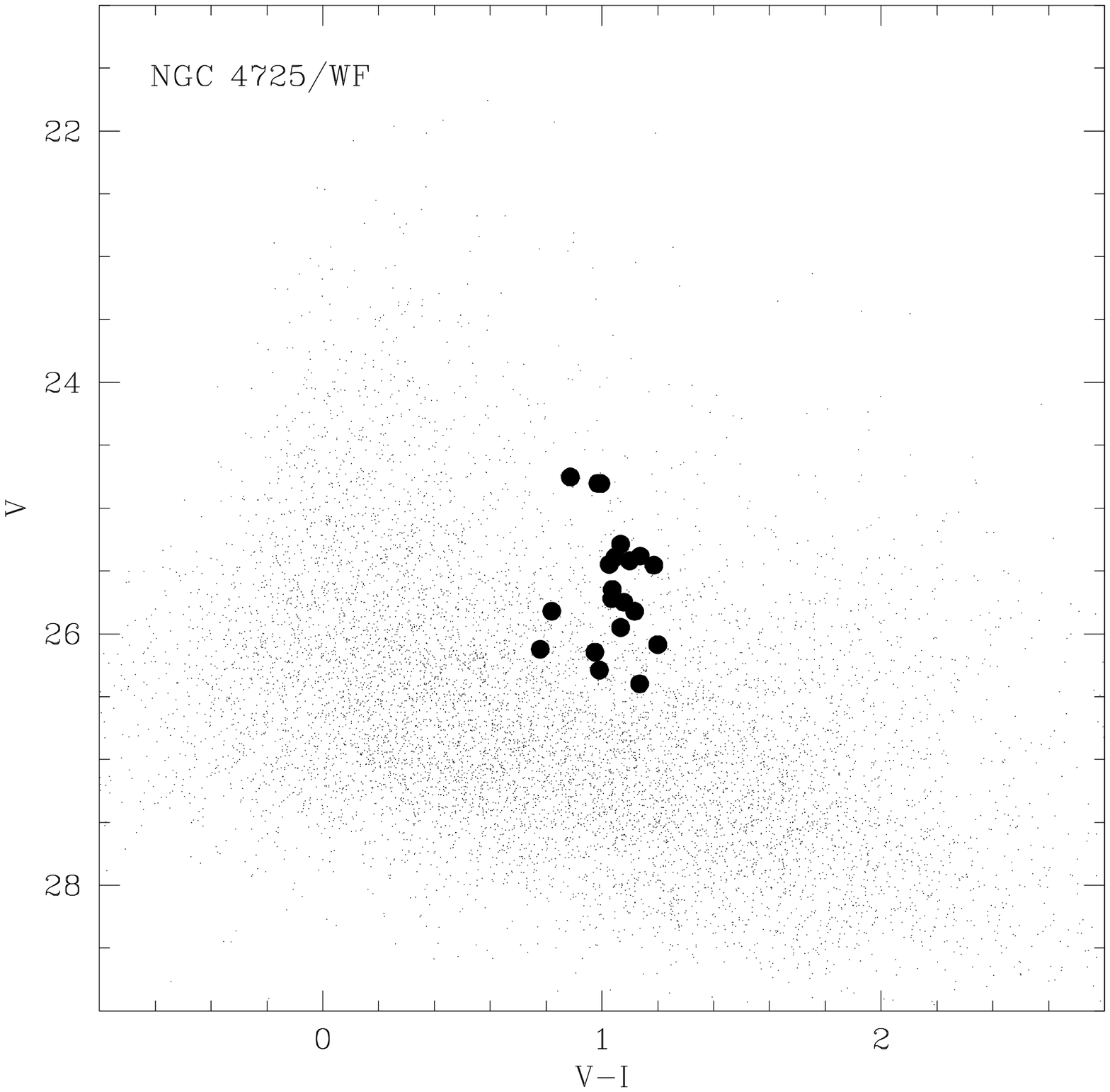]{
Calibrated ALLFRAME photometry (V,V$-$I) color-magnitude diagram for the three
WFC chips.
The filled circles represent the 20 NGC~4725 Cepheid
candidates of Table \ref{tbl:cepheids}.
\label{fig:cmd}}

\figcaption[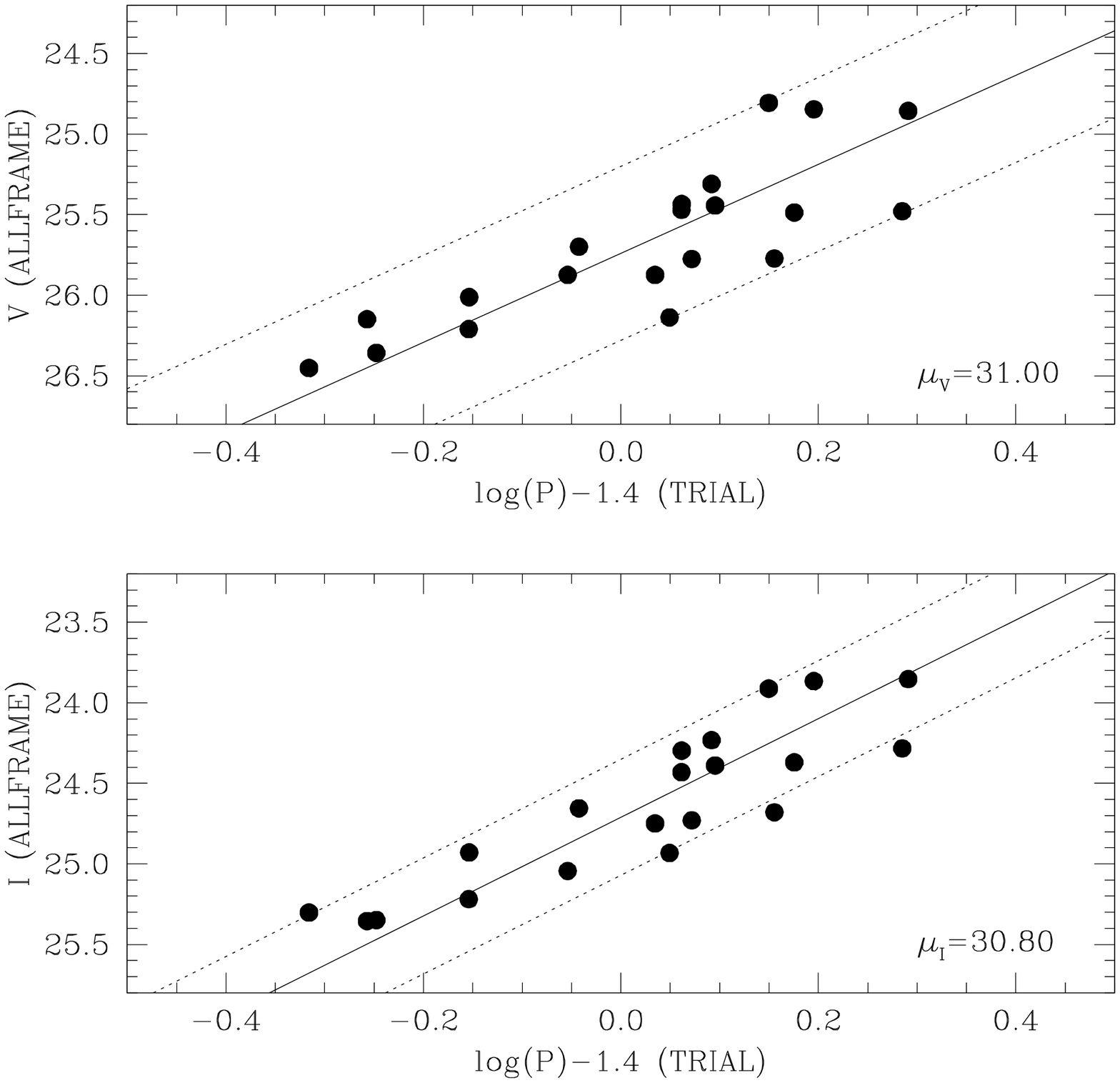]{
Period-luminosity relations in the V (top panel) and I (bottom panel) bands,
based on the calibrated ALLFRAME photometry.  The filled circles represent the
20 high-quality NGC~4725 Cepheid candidates found by TRIAL (see Tables
\ref{tbl:cepheids} and \ref{tbl:cepheids2}).
The solid lines are least
squares fits, with the slope fixed to be that of the Madore \& Freedman (1991)
LMC PL-relations, while 
the dotted lines represent their corresponding 2$\sigma$ dispersion.
The inferred apparent distance moduli are then
$\mu_{\rm V}=31.00\pm 0.06$ mag (internal) and $\mu_{\rm I}=30.80\pm 0.06$ mag
(internal).
\label{fig:pl_all}}

\figcaption[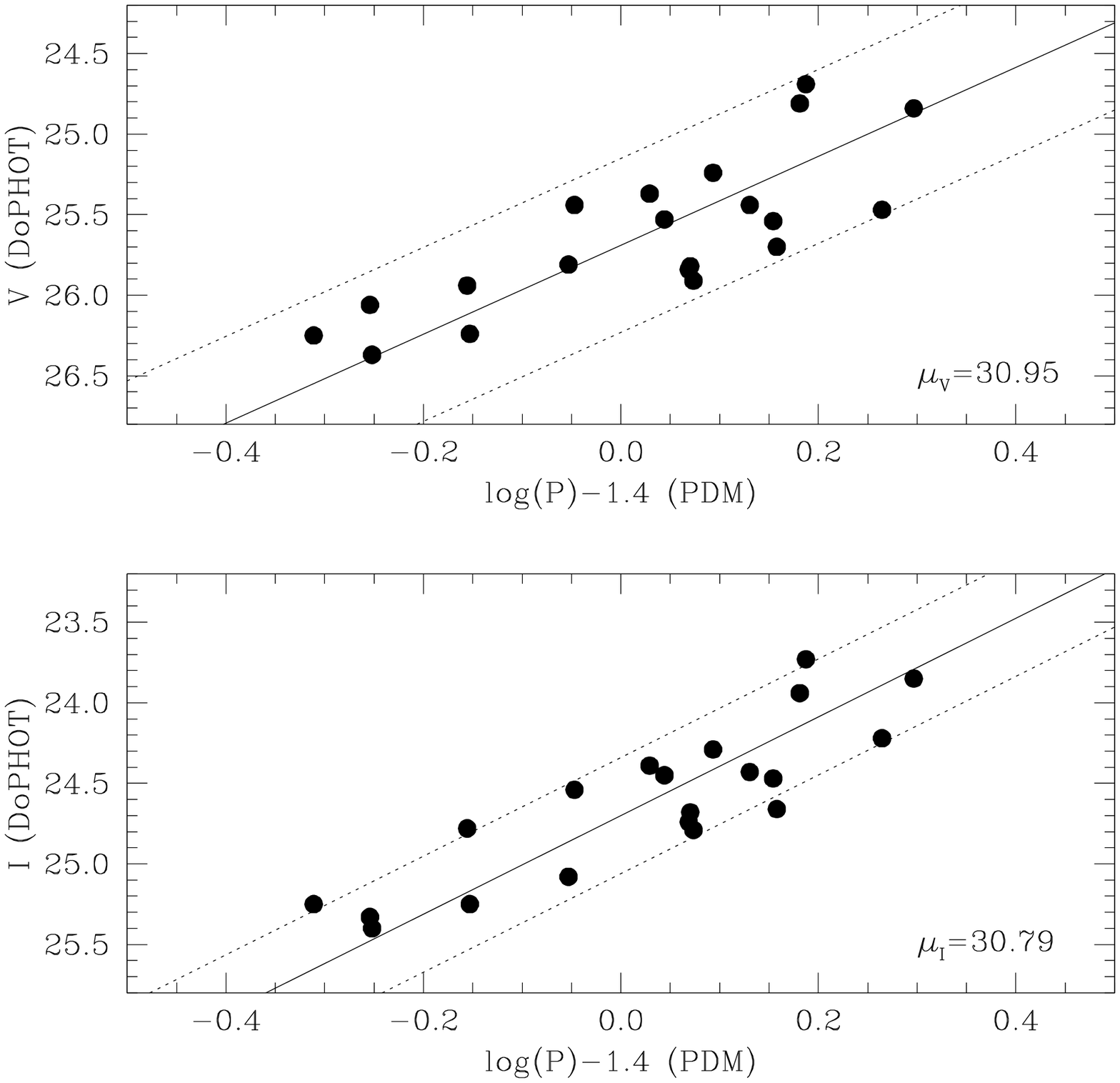]{
Period-luminosity relations in the V (top panel) and I (bottom panel) bands,
based on the calibrated DoPHOT photometry.  The filled circles represent the
20 high-quality NGC~4725 Cepheid candidates found by PDM (see Tables
\ref{tbl:cepheids} and \ref{tbl:cepheids2}).
The solid lines are least
squares fits, with the slope fixed to be that of the Madore \& Freedman (1991)
LMC PL-relations, while 
the dotted lines represent their corresponding 2$\sigma$ dispersion.
The inferred apparent distance moduli are then
$\mu_{\rm V}=30.91\pm 0.07$ mag (internal) and $\mu_{\rm I}=30.76\pm 0.06$ mag
(internal).
\label{fig:pl_do}}

\figcaption[irtf.eps]{
IRTF absolute calibration, with 
absolute H$_{-0.5}$ magnitudes versus HI linewidth
$\log(\Delta{\rm V})$, from Aaronson \etal (1982) and Tormen \& Burstein 
(1995), respectively.  The interim calibration (represented by the solid
curve) parallels that of Mould \etal 
(1997), and is based upon the available sample of 19 calibrators with
Cepheid-derived distances.
The dashed curve is
Freedman's (1990) calibration, based upon 5 local calibrators.
The circled dot represents NGC~4725.
\label{fig:irtf}}

\clearpage


\begin{deluxetable}{ccrrcc}
\footnotesize
\tablecaption{HST Observations of NGC~4725
\label{tbl:observations}}
\tablewidth{0pt}
\tablehead{
\colhead{Epoch} & \colhead{Filename} & \colhead{Date} & \colhead{Julian
Date} & \colhead{Exposure Times (s)} & \colhead{Filter}
}
\startdata
  1 & u2782j01t/2t & 12/04/95 & 2449819.813 & 1500\quad\quad 1000 & F555W \nl
  2 & u2782k01t/2t & 21/04/95 & 2449828.528 & 1500\quad\quad 1000 & F555W \nl
  3 & u2782l01t/2t & 02/05/95 & 2449839.777 & 1500\quad\quad 1000 & F555W \nl
  4 & u2782m01t/2t & 05/05/95 & 2449842.722 & 1500\quad\quad 1000 & F555W \nl
  5 & u2782n01t/2t & 07/05/95 & 2449845.269 & 1500\quad\quad 1000 & F555W \nl
  6 & u2782o01t/2t & 11/05/95 & 2449848.756 & 1500\quad\quad 1000 & F555W \nl
  7 & u2782p01t/2t & 15/05/95 & 2449852.993 & 1500\quad\quad 1000 & F555W \nl
  8 & u2782q01t/2t & 19/05/95 & 2449856.946 & 1500\quad\quad 1000 & F555W \nl
  9 & u2782r01p/2p & 24/05/95 & 2449862.174 & 1500\quad\quad 1000 & F555W \nl
 10$\,\,\,$ & u2782s01p/2p & 30/05/95 & 2449868.206 & 1500\quad\quad 1000 & F555W \nl
 11$\,\,\,$ & u2782t01t/2t & 06/06/95 & 2449874.974 & 1500\quad\quad 1000 & F555W \nl
 12$\,\,\,$ & u2782u01t/2t & 14/06/95 & 2449883.417 & 1500\quad\quad 1000 & F555W \nl
 13$\,\,\,$ & u2s76001t/2t & 29/04/96 & 2450203.095 & 1100\quad\quad 1100 & F555W \nl\nl
  2 & u2782k03t/4t & 21/04/95 & 2449828.593 & 1000\quad\quad 1500 & F814W \nl
  3 & u2782l03t/4t & 02/05/95 & 2449839.850 & 1000\quad\quad 1500 & F814W \nl
  8 & u2782q03t/4t & 19/05/95 & 2449857.013 & 1000\quad\quad 1500 & F814W \nl
 12$\,\,\,$ & u2782u03t/4t & 14/06/95 & 2449883.482 & 1000\quad\quad 1500 & F814W \nl\nl
  3 & u2782l05t/6t & 02/05/95 & 2449839.973 & 1500\quad\quad 1000 & F439W \nl
  8 & u2782q08t/9t & 19/05/95 & 2449857.149 & 1300\quad\quad 1200 & F439W \nl
\enddata
\end{deluxetable}

\clearpage

\begin{deluxetable}{lcrrccrrrr}
\footnotesize
\tablecaption{Reference Star Photometry.
\label{tbl:refstars}}
\tablewidth{0pt}
\tablehead{
\colhead{ID} & \colhead{Chip} & 
\colhead{X} & \colhead{Y} & \colhead{RA} & \colhead{Dec} & 
\multicolumn{2}{c}{ALLFRAME} & \multicolumn{2}{c}{DoPHOT} \\
\colhead{} & \colhead{} & \colhead{} & \colhead{} & \multicolumn{2}{c}{(J2000)}
& \colhead{V} & \colhead{I} &
\colhead{V} & \colhead{I}
}
\startdata
R01 & 2 & 721.2 & 210.7 & 12:50:35.44 & 25:31:41.8 & 24.23$\,\pm\,$0.01 & 22.40$\,\pm\,$0.01 & 24.11$\,\pm\,$0.02 & 22.42$\,\pm\,$0.02 \nl
R02 & 2 & 107.7 & 214.5 & 12:50:35.78 & 25:32:42.7 & 24.08$\,\pm\,$0.01 & 22.11$\,\pm\,$0.01 & 24.01$\,\pm\,$0.02 & 22.03$\,\pm\,$0.01 \nl
R03 & 2 & 451.8 & 258.5 & 12:50:35.25 & 25:32:08.9 & 23.97$\,\pm\,$0.01 & 23.94$\,\pm\,$0.03 & 23.95$\,\pm\,$0.02 & 23.86$\,\pm\,$0.03 \nl
R04 & 2 & 684.3 & 278.6 & 12:50:34.96 & 25:31:46.0 & 23.97$\,\pm\,$0.01 & 23.76$\,\pm\,$0.03 & 24.01$\,\pm\,$0.02 & 23.79$\,\pm\,$0.02 \nl
R05 & 2 & 155.4 & 291.7 & 12:50:35.18 & 25:32:38.6 & 24.24$\,\pm\,$0.01 & 23.89$\,\pm\,$0.03 & 24.20$\,\pm\,$0.02 & 23.84$\,\pm\,$0.05 \nl
R06 & 2 & 150.3 & 322.9 & 12:50:34.96 & 25:32:39.3 & 23.53$\,\pm\,$0.01 & 23.19$\,\pm\,$0.02 & 23.50$\,\pm\,$0.01 & 23.20$\,\pm\,$0.05 \nl
R07 & 2 & 791.9 & 344.7 & 12:50:34.41 & 25 31:35.9 & 23.09$\,\pm\,$0.01 & 22.82$\,\pm\,$0.02 & 23.13$\,\pm\,$0.02 & 22.87$\,\pm\,$0.03\nl
R08 & 2 & 613.1 & 351.9 & 12:50:34.47 & 25:31:53.7 & 23.02$\,\pm\,$0.01 & 21.90$\,\pm\,$0.01 & 23.08$\,\pm\,$0.01 & 21.95$\,\pm\,$0.02 \nl
R09 & 2 & 631.3 & 360.8 & 12:50:34.39 & 25:31:51.9 & 22.78$\,\pm\,$0.02 & 22.55$\,\pm\,$0.01 & 22.89$\,\pm\,$0.01 & 22.57$\,\pm\,$0.01 \nl
R10 & 2 & 414.3 & 463.4 & 12:50:33.77 & 25:32:14.3 & 23.07$\,\pm\,$0.01 & 22.94$\,\pm\,$0.02 & 23.03$\,\pm\,$0.03 & 22.98$\,\pm\,$0.02 \nl
R11 & 2 & 105.5 & 533.9 & 12:50:33.44 & 25:32:45.5 & 24.12$\,\pm\,$0.01 & 24.12$\,\pm\,$0.03 & 23.92$\,\pm\,$0.03 & 24.03$\,\pm\,$0.05 \nl
R12 & 2 &  84.2 & 605.4 & 12:50:32.92 & 25:32:48.2 & 22.42$\,\pm\,$0.01 & 22.06$\,\pm\,$0.01 & 22.44$\,\pm\,$0.02 & 21.97$\,\pm\,$0.06 \nl
R13 & 2 &  81.5 & 624.7 & 12:50:32.78 & 25:32:48.6 & 23.34$\,\pm\,$0.01 & 23.06$\,\pm\,$0.02 & 23.36$\,\pm\,$0.02 & 23.06$\,\pm\,$0.03 \nl
R14 & 2 & 178.6 & 653.5 & 12:50:32.51 & 25:32:39.2 & 23.52$\,\pm\,$0.01 & 23.02$\,\pm\,$0.02 & 23.46$\,\pm\,$0.02 & 22.98$\,\pm\,$0.01 \nl
R15 & 2 & 377.2 & 673.5 & 12:50:32.25 & 25:32:19.7 & 24.51$\,\pm\,$0.02 & 23.44$\,\pm\,$0.04 & 24.54$\,\pm\,$0.02 & 23.41$\,\pm\,$0.03 \nl
R16 & 2 & 193.4 & 680.6 & 12:50:32.31 & 25:32:38.0 & 24.17$\,\pm\,$0.01 & 21.61$\,\pm\,$0.01 & 24.06$\,\pm\,$0.03 & 21.54$\,\pm\,$0.03 \nl
R17 & 2 & 344.4 & 680.7 & 12:50:32.22 & 25:32:23.0 & 24.41$\,\pm\,$0.01 & 24.37$\,\pm\,$0.05 & 24.40$\,\pm\,$0.02 & 24.39$\,\pm\,$0.03 \nl
 \tablevspace{6pt}
R18 & 3 & 378.9 & 225.0 & 12:50:34.77 & 25:33:07.4 & 23.44$\,\pm\,$0.01 & 22.95$\,\pm\,$0.02 & 23.53$\,\pm\,$0.03 & 23.05$\,\pm\,$0.02 \nl
R19 & 3 & 445.5 & 240.4 & 12:50:34.29 & 25:33:09.5 & 23.87$\,\pm\,$0.01 & 23.83$\,\pm\,$0.03 & 23.92$\,\pm\,$0.03 & 23.84$\,\pm\,$0.02 \nl
R20 & 3 & 445.1 & 269.2 & 12:50:34.31 & 25:33:12.3 & 23.48$\,\pm\,$0.01 & 23.44$\,\pm\,$0.02 & 23.58$\,\pm\,$0.07 & 23.46$\,\pm\,$0.01 \nl
R21 & 3 & 642.8 & 272.0 & 12:50:32.87 & 25:33:14.3 & 23.82$\,\pm\,$0.01 & 23.67$\,\pm\,$0.03 & 23.88$\,\pm\,$0.03 & 23.70$\,\pm\,$0.03 \nl
R22 & 3 & 184.4 & 286.8 & 12:50:36.24 & 25:33:11.8 & 24.23$\,\pm\,$0.01 & 22.83$\,\pm\,$0.02 & 24.28$\,\pm\,$0.02 & 22.84$\,\pm\,$0.03 \nl
R23 & 3 & 648.7 & 354.7 & 12:50:32.88 & 25:33:22.6 & 22.71$\,\pm\,$0.01 & 22.42$\,\pm\,$0.01 & 22.80$\,\pm\,$0.01 & 22.51$\,\pm\,$0.02 \nl
R24 & 3 & 265.5 & 493.9 & 12:50:35.77 & 25:33:33.1 & 24.10$\,\pm\,$0.01 & 23.86$\,\pm\,$0.03 & 24.12$\,\pm\,$0.03 & 23.88$\,\pm\,$0.02 \nl
R25 & 3 & 528.1 & 500.0 & 12:50:33.85 & 25:33:35.9 & 24.36$\,\pm\,$0.01 & 23.96$\,\pm\,$0.04 & 24.41$\,\pm\,$0.02 & 24.05$\,\pm\,$0.01 \nl
R26 & 3 & 456.9 & 544.3 & 12:50:34.40 & 25:33:39.7 & 22.63$\,\pm\,$0.01 & 22.10$\,\pm\,$0.01 & 22.68$\,\pm\,$0.01 & 22.14$\,\pm\,$0.03 \nl
R27 & 3 & 394.3 & 547.2 & 12:50:34.86 & 25:33:39.5 & 24.08$\,\pm\,$0.01 & 24.07$\,\pm\,$0.03 & 24.11$\,\pm\,$0.01 & 24.00$\,\pm\,$0.04 \nl
R28 & 3 & 473.7 & 568.4 & 12:50:34.30 & 25:33:42.2 & 23.44$\,\pm\,$0.01 & 23.50$\,\pm\,$0.02 & 23.48$\,\pm\,$0.02 & 23.52$\,\pm\,$0.02 \nl
R29 & 3 & 602.5 & 697.2 & 12:50:33.43 & 25:33:56.1 & 23.57$\,\pm\,$0.01 & 23.31$\,\pm\,$0.02 & 23.58$\,\pm\,$0.01 & 23.32$\,\pm\,$0.01 \nl
R30 & 3 & 494.7 & 767.5 & 12:50:34.27 & 25:34:02.2 & 24.00$\,\pm\,$0.01 & 23.93$\,\pm\,$0.03 & 23.95$\,\pm\,$0.04 & 23.95$\,\pm\,$0.02 \nl
 \tablevspace{6pt}
R31 & 4 & 210.1 & 148.3 & 12:50:38.04 & 25:33:02.1 & 24.23$\,\pm\,$0.01 & 22.75$\,\pm\,$0.04 & 24.22$\,\pm\,$0.01 & 22.78$\,\pm\,$0.03 \nl
R32 & 4 & 386.9 & 229.9 & 12:50:38.77 & 25:33:18.9 & 22.40$\,\pm\,$0.01 & 22.39$\,\pm\,$0.01 & 22.43$\,\pm\,$0.02 & 22.48$\,\pm\,$0.02 \nl
R33 & 4 & 260.8 & 239.2 & 12:50:38.75 & 25:33:06.3 & 22.92$\,\pm\,$0.01 & 22.04$\,\pm\,$0.03 & 22.95$\,\pm\,$0.02 & 22.09$\,\pm\,$0.04 \nl
R34 & 4 & 394.1 & 278.2 & 12:50:39.13 & 25:33:19.1 & 24.39$\,\pm\,$0.01 & 24.08$\,\pm\,$0.03 & 24.39$\,\pm\,$0.02 & 24.11$\,\pm\,$0.02 \nl
R35 & 4 & 645.0 & 311.6 & 12:50:39.55 & 25:33:43.7 & 24.76$\,\pm\,$0.01 & 24.07$\,\pm\,$0.21 & 24.71$\,\pm\,$0.02 & 24.33$\,\pm\,$0.02 \nl
R36 & 4 & 796.6 & 348.8 & 12:50:39.93 & 25:33:58.3 & 23.97$\,\pm\,$0.01 & 23.07$\,\pm\,$0.02 & 23.85$\,\pm\,$0.02 & 23.03$\,\pm\,$0.02 \nl
R37 & 4 & 636.5 & 373.0 & 12:50:39.99 & 25:33:42.2 & 24.07$\,\pm\,$0.01 & 23.95$\,\pm\,$0.04 & 24.15$\,\pm\,$0.01 & 24.07$\,\pm\,$0.06 \nl
R38 & 4 & 498.9 & 539.2 & 12:50:41.12 & 25:33:27.0 & 23.72$\,\pm\,$0.02 & 23.66$\,\pm\,$0.06 & 23.69$\,\pm\,$0.02 & 23.74$\,\pm\,$0.02 \nl
R39 & 4 & 162.8 & 525.8 & 12:50:40.78 & 25:32:53.8 & 23.77$\,\pm\,$0.01 & 23.60$\,\pm\,$0.02 & 23.84$\,\pm\,$0.03 & 23.68$\,\pm\,$0.03 \nl
R40 & 4 & 202.9 & 544.6 & 12:50:40.94 & 25:32:57.6 & 20.14$\,\pm\,$0.01 & 17.14$\,\pm\,$0.02 & 20.29$\,\pm\,$0.05 & 17.13$\,\pm\,$0.03 \nl
R41 & 4 & 365.6 & 546.0 & 12:50:41.07 & 25:33:13.7 & 24.33$\,\pm\,$0.01 & 23.98$\,\pm\,$0.04 & 24.41$\,\pm\,$0.02 & 24.08$\,\pm\,$0.05 \nl
R42 & 4 & 287.3 & 558.0 & 12:50:41.10 & 25:33:05.8 & 23.50$\,\pm\,$0.01 & 23.06$\,\pm\,$0.02 & 23.56$\,\pm\,$0.02 & 23.13$\,\pm\,$0.03 \nl
R43 & 4 & 296.4 & 665.3 & 12:50:41.90 & 25:33:05.7 & 24.46$\,\pm\,$0.01 & 24.27$\,\pm\,$0.04 & 24.50$\,\pm\,$0.03 & 24.36$\,\pm\,$0.08 \nl
R44 & 4 & 458.1 & 671.7 & 12:50:42.06 & 25:33:21.7 & 20.31$\,\pm\,$0.00 & 18.76$\,\pm\,$0.02 & 20.31$\,\pm\,$0.03 & 18.86$\,\pm\,$0.01 \nl
R45 & 4 & 305.0 & 699.0 & 12:50:42.15 & 25:33:06.2 & 22.62$\,\pm\,$0.01 & 22.23$\,\pm\,$0.02 & 22.59$\,\pm\,$0.04 & 22.31$\,\pm\,$0.03 \nl
\enddata
\end{deluxetable}

\clearpage

\begin{deluxetable}{crrrrr}
\footnotesize
\tablecaption{Comparison of ALLFRAME and DoPHOT Magnitudes
\label{tbl:all_do_comp}}
\tablewidth{0pt}
\tablehead{
\colhead{Chip} & \colhead{\# Stars} & \colhead{$\Delta$V\tablenotemark{a}} & 
\colhead{$\sigma_{\Delta{\rm V}}$} &
\colhead{$\Delta$I\tablenotemark{a}} & \colhead{$\sigma_{\Delta{\rm I}}$}
}
\startdata
\multicolumn{6}{c}{\it Reference Stars}\nl
 2  & 17\quad & $+$0.022 & 0.073 & $+$0.017 & 0.050 \nl
 3  & 13\quad & $-$0.045 & 0.038 & $-$0.030 & 0.043 \nl
 4  & 15\quad & $-$0.020 & 0.063 & $-$0.075 & 0.065 \nl
2-4 & 45\quad & $-$0.011 & 0.068 & $-$0.027 & 0.066 \nl\nl
\multicolumn{6}{c}{\it Cepheids}\nl
 2  & 12\quad                  & $+$0.088 & 0.097 & $+$0.038 & 0.081 \nl
 3  &  3\quad                  & $-$0.030 & 0.016 & $-$0.060 & 0.029 \nl
 4  &  5\quad                  & $+$0.036 & 0.055 & $+$0.006 & 0.042 \nl
2-4 & 20\quad                  & $+$0.057 & 0.091 & $+$0.016 & 0.075 \nl
\enddata
\tablenotetext{a}{$\Delta\equiv$ALLFRAME-DoPHOT.}
\end{deluxetable}

\clearpage

\begin{deluxetable}{ccrrcc}
\footnotesize
\tablecaption{Cepheid Candidates Detected in NGC~4725 - Coordinates
\label{tbl:cepheids}}
\tablewidth{0pt}
\tablehead{
\colhead{ID} & \colhead{Chip} & \colhead{X} & \colhead{Y} & \colhead{RA} &
\colhead{Dec} \\
\colhead{} & \colhead{} & \colhead{} & \colhead{} & \multicolumn{2}{c}{(J2000)}
}
\startdata
C01 & 2 & 594.0 & 100.2 & 12:50:36.33 & 25:31:53.5 \nl
C02 & 2 & 570.6 & 226.2 & 12:50:35.42 & 25:31:56.9 \nl
C03 & 2 & 523.0 & 242.7 & 12:50:35.32 & 25:32:01.7 \nl
C04 & 2 & 629.5 & 336.9 & 12:50:34.57 & 25:31:51.9 \nl
C05 & 2 & 558.8 & 338.6 & 12:50:34.60 & 25:31:58.9 \nl
C06 & 2 & 675.2 & 358.3 & 12:50:34.38 & 25:31:47.5 \nl
C07 & 2 & 160.9 & 473.7 & 12:50:33.85 & 25:32:39.5 \nl
C08 & 2 &  90.3 & 521.5 & 12:50:33.54 & 25:32:46.9 \nl
C09 & 2 & 183.7 & 570.6 & 12:50:33.12 & 25:32:38.0 \nl
C10 & 2 & 566.3 & 585.3 & 12:50:32.78 & 25:32:00.2 \nl
C11 & 2 & 465.9 & 593.0 & 12:50:32.79 & 25:32:10.2 \nl
C12 & 2 &  97.3 & 655.6 & 12:50:32.55 & 25:32:47.3 \nl\nl
C13 & 3 &  98.0 & 230.5 & 12:50:36.83 & 25:33:05.5 \nl
C14 & 3 & 353.9 & 420.4 & 12:50:35.08 & 25:33:26.5 \nl
C15 & 3 & 674.9 & 475.8 & 12:50:32.76 & 25:33:34.8 \nl\nl
C16 & 4 & 133.9 & 230.6 & 12:50:38.59 & 25:32:53.8 \nl
C17 & 4 & 687.7 & 268.9 & 12:50:39.27 & 25:33:48.3 \nl
C18 & 4 & 724.1 & 287.8 & 12:50:39.43 & 25:33:51.7 \nl
C19 & 4 & 490.8 & 333.6 & 12:50:39.60 & 25:33:28.2 \nl
C20 & 4 & 705.9 & 408.3 & 12:50:40.30 & 25:33:48.8 \nl
\enddata
\end{deluxetable}

\clearpage

\begin{deluxetable}{ccccccc}
\footnotesize
\tablecaption{Cepheids Detected in NGC~4725 - Properties\tablenotemark{a}
\label{tbl:cepheids2}}
\tablewidth{0pt}
\tablehead{
\colhead{ID} & \multicolumn{3}{c}{ALLFRAME/TRIAL} & \multicolumn{3}{c}{DoPHOT/PDM} \\
\colhead{} & \colhead{Period (d)} & \colhead{V} & \colhead{I} & \colhead{Period (d)} & \colhead{V} & \colhead{I}
}
\startdata
C01                  & $28.95\pm 0.05$ & $25.43\pm 0.03$ & $24.30\pm 0.06$
	& 26.9          & $25.37\pm 0.03$ & $24.39\pm 0.13$ \nl
C02                  & $12.14\pm 0.02$ & $26.45\pm 0.04$ & $25.30\pm 0.07$
	& 12.3          & $26.25\pm 0.04$ & $25.25\pm 0.07$ \nl
C03                  & $17.63\pm 0.04$ & $26.01\pm 0.03$ & $24.93\pm 0.05$
	& 17.6          & $25.94\pm 0.04$ & $24.78\pm 0.06$ \nl
C04                  & $22.19\pm 0.09$ & $25.87\pm 0.03$ & $25.04\pm 0.04$
	& 22.2          & $25.81\pm 0.03$ & $25.08\pm 0.07$ \nl
C05                  & $28.13\pm 0.28$ & $26.14\pm 0.04$ & $24.93\pm 0.06$
	& 29.8          & $25.91\pm 0.04$ & $24.79\pm 0.05$ \nl
C06                  & $49.09\pm 0.25$ & $24.86\pm 0.02$ & $23.85\pm 0.03$
	& 49.7          & $24.84\pm 0.02$ & $23.85\pm 0.05$ \nl
C07                  & $29.63\pm 0.08$ & $25.78\pm 0.02$ & $24.73\pm 0.05$
	& 29.4          & $25.84\pm 0.03$ & $24.74\pm 0.07$ \nl
C08                  & $31.29\pm 0.45$ & $25.44\pm 0.03$ & $24.39\pm 0.04$
	& 33.9          & $25.44\pm 0.03$ & $24.43\pm 0.05$ \nl
C09                  & $39.39\pm 0.06$ & $24.85\pm 0.01$ & $23.87\pm 0.02$
	& 38.7          & $24.69\pm 0.01$ & $23.73\pm 0.03$ \nl
C10                  & $35.46\pm 0.43$ & $24.81\pm 0.02$ & $23.91\pm 0.04$
	& 38.1          & $24.81\pm 0.02$ & $23.94\pm 0.05$ \nl
C11                  & $22.78\pm 0.02$ & $25.70\pm 0.04$ & $24.66\pm 0.05$
	& 22.5          & $25.44\pm 0.03$ & $24.54\pm 0.05$ \nl
C12                  & $27.20\pm 0.11$ & $25.87\pm 0.04$ & $24.75\pm 0.06$
	& 29.5          & $25.82\pm 0.03$ & $24.68\pm 0.05$ \nl
C13                  & $37.63\pm 0.17$ & $25.49\pm 0.02$ & $24.37\pm 0.03$
	& 35.8          & $25.54\pm 0.03$ & $24.47\pm 0.04$ \nl
C14                  & $17.62\pm 0.15$ & $26.21\pm 0.04$ & $25.22\pm 0.05$
	& 17.7          & $26.24\pm 0.04$ & $25.25\pm 0.07$ \nl
C15                  & $14.20\pm 0.03$ & $26.36\pm 0.03$ & $25.35\pm 0.05$
	& 14.1          & $26.37\pm 0.04$ & $25.40\pm 0.08$ \nl
C16                  & $35.93\pm 0.40$ & $25.77\pm 0.02$ & $24.68\pm 0.04$
	& 36.1          & $25.70\pm 0.03$ & $24.66\pm 0.06$ \nl
C17                  & $31.03\pm 0.12$ & $25.31\pm 0.02$ & $24.23\pm 0.03$
	& 31.1          & $25.24\pm 0.02$ & $24.29\pm 0.05$ \nl
C18                  & $28.93\pm 0.19$ & $25.47\pm 0.02$ & $24.43\pm 0.04$
	& 27.8          & $25.53\pm 0.02$ & $24.45\pm 0.05$ \nl
C19                  & $48.41\pm 0.44$ & $25.48\pm 0.02$ & $24.28\pm 0.03$
	& 46.2          & $25.47\pm 0.03$ & $24.22\pm 0.03$ \nl
C20                  & $13.90\pm 0.03$ & $26.15\pm 0.04$ & $25.36\pm 0.06$
	& 14.0          & $26.06\pm 0.04$ & $25.33\pm 0.07$ \nl
\enddata
\tablenotetext{a}{Notes:
C01-bright,isolated.  C02-faint,isolated.  C03-isolated.  C04-isolated.
C05-bright,neighbor at $\sim 0^{\prime\prime}.27$.  C06-bright,near dust lane.
C07-bright,isolated.  C08-bright,isolated.  C09-bright,isolated.
C10-crowded,bright background.  C11-isolated,bright background.  C12-isolated.
C13-bright,isolated.  C14-faint,isolated.  C15-faint,isolated.  C16-isolated.
C17-bright,isolated.  C18-isolated.  C19-isolated.  C20-isolated.}
\end{deluxetable}

\clearpage

\begin{deluxetable}{cccccccc}
\footnotesize
\tablecaption{Measured ALLFRAME
Magnitudes and Standard Errors
\label{tbl:magnitudes}}
\tablewidth{0pt}
\tablehead{
\colhead{HJD} & \colhead{Filter} & \colhead{magnitude} & \colhead{magnitude} & \colhead{magnitude} & \colhead{magnitude} & \colhead{magnitude} & \colhead{magnitude} 
}
\startdata
              &   &        C01        &        C02        &        C03        &        C04        &        C05        &        C06        \nl
  2449819.813 & V & 26.16 $\pm$ 0.18  & 27.00 $\pm$ 0.40  & 25.90 $\pm$ 0.11  & 25.68 $\pm$ 0.12  & 26.38 $\pm$ 0.22  & 24.95 $\pm$ 0.12  \nl
  2449819.867 & V & 26.21 $\pm$ 0.17  & 26.42 $\pm$ 0.31  & 26.05 $\pm$ 0.19  & 25.61 $\pm$ 0.17  & 26.72 $\pm$ 0.37  & 24.99 $\pm$ 0.08  \nl
  2449828.528 & V & 25.07 $\pm$ 0.05  & 25.65 $\pm$ 1.03  & 26.53 $\pm$ 0.19  & 26.50 $\pm$ 0.23  & 25.76 $\pm$ 0.06  & 25.18 $\pm$ 0.12  \nl
  2449828.579 & V & 25.22 $\pm$ 0.14  & 25.77 $\pm$ 0.14  & 26.70 $\pm$ 0.38  & 26.29 $\pm$ 0.16  & 25.55 $\pm$ 0.13  & 25.13 $\pm$ 0.15  \nl
  2449828.593 & I & 23.65 $\pm$ 0.41  & 25.04 $\pm$ 0.18  & 25.34 $\pm$ 0.30  & 25.26 $\pm$ 0.26  & 24.74 $\pm$ 0.16  & 24.11 $\pm$ 0.15  \nl
  2449828.649 & I & 24.17 $\pm$ 0.24  & 25.12 $\pm$ 0.21  & 25.06 $\pm$ 0.19  & 25.37 $\pm$ 0.20  & 24.85 $\pm$ 0.14  & 23.95 $\pm$ 0.11  \nl
  2449839.777 & V & 25.67 $\pm$ 0.09  & 26.45 $\pm$ 0.16  & 25.95 $\pm$ 0.12  & 25.60 $\pm$ 0.11  & 26.18 $\pm$ 0.20  & 23.65 $\pm$ 0.21  \nl
  2449839.836 & V & 25.69 $\pm$ 0.19  & 26.41 $\pm$ 0.16  & 26.30 $\pm$ 0.22  & 25.44 $\pm$ 0.11  & 26.27 $\pm$ 0.39  & 25.07 $\pm$ 0.11  \nl
  2449839.850 & I & 24.38 $\pm$ 0.10  & 25.95 $\pm$ 0.43  & 24.81 $\pm$ 0.17  & 24.82 $\pm$ 0.13  & 25.39 $\pm$ 0.31  & 23.81 $\pm$ 0.06  \nl
  2449839.907 & I & 24.77 $\pm$ 0.16  & 25.17 $\pm$ 0.20  & 24.96 $\pm$ 0.13  & 24.98 $\pm$ 0.12  & 25.18 $\pm$ 0.27  & 24.07 $\pm$ 0.13  \nl
  2449842.722 & V & 24.33 $\pm$ 0.44  & 26.17 $\pm$ 0.16  & 26.46 $\pm$ 0.21  & 25.61 $\pm$ 0.14  & 26.66 $\pm$ 0.27  & 24.81 $\pm$ 0.10  \nl
  2449842.785 & V & 26.00 $\pm$ 0.14  & 26.25 $\pm$ 0.22  & 26.71 $\pm$ 0.35  & 25.88 $\pm$ 0.19  & 26.26 $\pm$ 0.23  & 24.82 $\pm$ 0.10  \nl
  2449845.269 & V & 25.98 $\pm$ 0.17  & 26.68 $\pm$ 0.18  & 26.78 $\pm$ 0.22  & 26.05 $\pm$ 0.13  & 26.91 $\pm$ 0.30  & 24.41 $\pm$ 0.06  \nl
  2449845.288 & V & 26.09 $\pm$ 0.20  & 26.77 $\pm$ 0.40  & 26.48 $\pm$ 0.32  & 26.05 $\pm$ 0.21  & 26.65 $\pm$ 0.35  & 24.59 $\pm$ 0.07  \nl
  2449848.756 & V & 25.82 $\pm$ 0.08  & 27.23 $\pm$ 0.33  & 25.99 $\pm$ 0.12  & 26.06 $\pm$ 0.14  & 26.68 $\pm$ 0.32  & 24.43 $\pm$ 0.05  \nl
  2449848.819 & V & 25.91 $\pm$ 0.15  & 27.34 $\pm$ 0.47  & 25.91 $\pm$ 0.17  & 26.01 $\pm$ 0.52  & 26.44 $\pm$ 0.33  & 24.43 $\pm$ 0.06  \nl
  2449852.993 & V & 24.79 $\pm$ 0.07  & 25.74 $\pm$ 0.10  & 25.90 $\pm$ 0.10  & 24.10 $\pm$ 0.22  & 26.00 $\pm$ 0.13  & 24.62 $\pm$ 0.07  \nl
  2449853.044 & V & 24.85 $\pm$ 0.09  & 23.29 $\pm$ 0.46  & 24.33 $\pm$ 0.65  & 26.74 $\pm$ 0.38  & 26.32 $\pm$ 0.27  & 23.58 $\pm$ 0.35  \nl
  2449856.946 & V & 25.07 $\pm$ 0.07  & 26.98 $\pm$ 0.34  & 26.05 $\pm$ 0.18  & 25.82 $\pm$ 0.12  & 25.83 $\pm$ 0.15  & 24.67 $\pm$ 0.08  \nl
  2449856.999 & V & 25.04 $\pm$ 0.10  & 26.37 $\pm$ 0.20  & 25.95 $\pm$ 0.23  & 25.82 $\pm$ 0.19  & 25.44 $\pm$ 0.12  & 24.81 $\pm$ 0.10  \nl
  2449857.013 & I & 23.33 $\pm$ 0.81  & 25.41 $\pm$ 0.21  & 24.73 $\pm$ 0.12  & 24.82 $\pm$ 0.17  & 24.57 $\pm$ 0.19  & 23.72 $\pm$ 0.08  \nl
  2449857.069 & I & 23.47 $\pm$ 0.46  & 25.26 $\pm$ 0.17  & 24.94 $\pm$ 0.17  & 25.15 $\pm$ 0.12  & 24.83 $\pm$ 0.13  & 23.76 $\pm$ 0.09  \nl
  2449857.082 & V & 25.49 $\pm$ 0.25  & 25.96 $\pm$ 0.30  & 25.95 $\pm$ 0.40  & 26.59 $\pm$ 0.75  & 25.99 $\pm$ 0.51  & 24.76 $\pm$ 0.19  \nl
  2449857.133 & I & 24.22 $\pm$ 0.21  & $\ldots$          & 24.65 $\pm$ 0.30  & 25.47 $\pm$ 0.75  & 24.35 $\pm$ 0.30  & 23.93 $\pm$ 0.18  \nl
  2449862.174 & V & 25.43 $\pm$ 0.10  & 27.06 $\pm$ 0.35  & 26.71 $\pm$ 0.22  & 25.62 $\pm$ 0.10  & 25.84 $\pm$ 0.12  & 24.82 $\pm$ 0.09  \nl
  2449862.233 & V & 24.98 $\pm$ 0.64  & 27.19 $\pm$ 0.50  & 26.42 $\pm$ 0.20  & 25.54 $\pm$ 0.10  & 26.09 $\pm$ 0.13  & 24.74 $\pm$ 0.10  \nl
  2449868.206 & V & 25.80 $\pm$ 0.13  & 26.56 $\pm$ 0.21  & 25.38 $\pm$ 0.10  & 26.09 $\pm$ 0.12  & 25.48 $\pm$ 0.22  & 24.93 $\pm$ 0.08  \nl
  2449868.264 & V & 25.78 $\pm$ 0.16  & 26.38 $\pm$ 0.23  & 25.51 $\pm$ 0.09  & 25.99 $\pm$ 0.19  & 25.58 $\pm$ 0.16  & 24.98 $\pm$ 0.09  \nl
  2449874.974 & V & 26.06 $\pm$ 0.15  & 26.72 $\pm$ 0.18  & 25.78 $\pm$ 0.10  & 26.65 $\pm$ 0.33  & 26.67 $\pm$ 0.28  & 25.10 $\pm$ 0.06  \nl
  2449875.025 & V & 26.08 $\pm$ 0.22  & 26.97 $\pm$ 0.48  & 26.24 $\pm$ 0.35  & 26.63 $\pm$ 0.30  & 26.67 $\pm$ 0.34  & 25.10 $\pm$ 0.10  \nl
  2449883.417 & V & 25.00 $\pm$ 0.07  & 26.71 $\pm$ 0.17  & 26.05 $\pm$ 0.14  & 25.47 $\pm$ 0.07  & 25.73 $\pm$ 0.11  & 25.34 $\pm$ 0.11  \nl
  2449883.468 & V & 24.87 $\pm$ 0.07  & 26.84 $\pm$ 0.27  & 26.41 $\pm$ 0.25  & 25.78 $\pm$ 0.20  & 25.79 $\pm$ 0.17  & 25.21 $\pm$ 0.12  \nl
  2449883.482 & I & 24.16 $\pm$ 0.09  & 25.17 $\pm$ 0.24  & 24.95 $\pm$ 0.21  & 24.78 $\pm$ 0.15  & 24.78 $\pm$ 0.16  & 24.25 $\pm$ 0.11  \nl
  2449883.538 & I & 23.86 $\pm$ 0.16  & 25.41 $\pm$ 0.22  & 25.30 $\pm$ 0.15  & 24.86 $\pm$ 0.10  & 24.54 $\pm$ 0.12  & 24.06 $\pm$ 0.28  \nl
  2450203.095 & V & 24.97 $\pm$ 0.09  & 26.70 $\pm$ 0.33  & 25.47 $\pm$ 0.10  & 26.12 $\pm$ 0.17  & 26.11 $\pm$ 0.29  & 24.75 $\pm$ 0.06  \nl
  2450203.109 & V & 24.93 $\pm$ 0.08  & 26.73 $\pm$ 0.27  & 25.42 $\pm$ 0.13  & 26.29 $\pm$ 0.16  & 26.20 $\pm$ 0.21  & 24.69 $\pm$ 0.06  \nl
              &   &        C07        &        C08        &        C09        &        C10        &        C11        &        C12        \nl
  2449819.813 & V & 26.18 $\pm$ 0.11  & 25.64 $\pm$ 0.12  & 24.80 $\pm$ 0.06  & 25.36 $\pm$ 0.11  & 25.30 $\pm$ 0.09  & 26.14 $\pm$ 0.17  \nl
  2449819.867 & V & 26.23 $\pm$ 0.22  & 25.56 $\pm$ 0.11  & 24.77 $\pm$ 0.06  & 25.44 $\pm$ 0.13  & 23.33 $\pm$ 0.27  & 25.99 $\pm$ 0.33  \nl
  2449828.528 & V & 26.51 $\pm$ 0.19  & 25.98 $\pm$ 0.14  & 25.12 $\pm$ 0.08  & 24.44 $\pm$ 0.06  & 26.01 $\pm$ 0.21  & 25.47 $\pm$ 0.14  \nl
  2449828.579 & V & 26.92 $\pm$ 0.39  & 26.27 $\pm$ 0.23  & 25.23 $\pm$ 0.10  & 24.33 $\pm$ 0.31  & 26.27 $\pm$ 0.51  & 25.54 $\pm$ 0.14  \nl
  2449828.593 & I & 25.22 $\pm$ 0.23  & 24.69 $\pm$ 0.16  & 23.97 $\pm$ 0.09  & 23.64 $\pm$ 0.13  & 24.83 $\pm$ 0.13  & 24.57 $\pm$ 0.11  \nl
  2449828.649 & I & 25.06 $\pm$ 0.30  & 24.90 $\pm$ 0.16  & 23.88 $\pm$ 0.08  & 23.57 $\pm$ 0.05  & 25.05 $\pm$ 0.18  & 24.56 $\pm$ 0.13  \nl
  2449839.777 & V & 25.66 $\pm$ 0.11  & 25.05 $\pm$ 0.08  & 25.16 $\pm$ 0.09  & 24.69 $\pm$ 0.08  & 25.03 $\pm$ 0.08  & 25.96 $\pm$ 0.17  \nl
  2449839.836 & V & 25.56 $\pm$ 0.13  & 25.13 $\pm$ 0.08  & 25.18 $\pm$ 0.08  & 24.91 $\pm$ 0.12  & 25.21 $\pm$ 0.10  & 26.41 $\pm$ 0.38  \nl
  2449839.850 & I & 24.50 $\pm$ 0.13  & 24.27 $\pm$ 0.10  & 24.09 $\pm$ 0.10  & 23.78 $\pm$ 0.10  & 24.54 $\pm$ 0.17  & 22.93 $\pm$ 0.34  \nl
  2449839.907 & I & 24.62 $\pm$ 0.11  & 24.14 $\pm$ 0.07  & 24.17 $\pm$ 0.11  & 23.79 $\pm$ 0.09  & 24.35 $\pm$ 0.09  & 24.82 $\pm$ 0.13  \nl
  2449842.722 & V & 25.69 $\pm$ 0.13  & 25.27 $\pm$ 0.08  & 24.78 $\pm$ 0.07  & 25.04 $\pm$ 0.10  & 24.39 $\pm$ 0.27  & 26.66 $\pm$ 0.21  \nl
  2449842.785 & V & 25.77 $\pm$ 0.14  & 25.31 $\pm$ 0.11  & 24.76 $\pm$ 0.09  & 24.99 $\pm$ 0.12  & 25.38 $\pm$ 0.14  & 26.43 $\pm$ 0.29  \nl
  2449845.269 & V & 25.70 $\pm$ 0.10  & 24.83 $\pm$ 0.33  & 24.38 $\pm$ 0.05  & 24.91 $\pm$ 0.07  & 25.54 $\pm$ 0.08  & 26.29 $\pm$ 0.22  \nl
  2449845.288 & V & 25.23 $\pm$ 1.11  & 25.51 $\pm$ 0.17  & 24.41 $\pm$ 0.06  & 24.88 $\pm$ 0.07  & 25.65 $\pm$ 0.13  & 26.36 $\pm$ 0.30  \nl
  2449848.756 & V & 26.19 $\pm$ 0.14  & 25.59 $\pm$ 0.10  & 24.44 $\pm$ 0.06  & 25.04 $\pm$ 0.10  & 25.95 $\pm$ 0.13  & 26.14 $\pm$ 0.17  \nl
  2449848.819 & V & 26.06 $\pm$ 0.17  & 25.57 $\pm$ 0.09  & 24.47 $\pm$ 0.08  & 24.91 $\pm$ 0.09  & 25.66 $\pm$ 0.13  & 26.17 $\pm$ 0.29  \nl
  2449852.993 & V & 26.64 $\pm$ 0.22  & 24.09 $\pm$ 0.26  & 24.61 $\pm$ 0.06  & 25.12 $\pm$ 0.11  & 25.39 $\pm$ 0.73  & 24.22 $\pm$ 0.60  \nl
  2449853.044 & V & 26.52 $\pm$ 0.15  & 25.77 $\pm$ 0.29  & 24.75 $\pm$ 0.08  & 25.35 $\pm$ 0.14  & 25.96 $\pm$ 0.17  & 25.47 $\pm$ 0.09  \nl
  2449856.946 & V & 26.27 $\pm$ 0.21  & 25.92 $\pm$ 0.22  & 24.82 $\pm$ 0.06  & 25.37 $\pm$ 0.12  & 26.28 $\pm$ 0.17  & 25.59 $\pm$ 0.16  \nl
  2449856.999 & V & 26.38 $\pm$ 0.26  & 26.24 $\pm$ 0.18  & 24.71 $\pm$ 0.09  & 25.36 $\pm$ 0.14  & 25.90 $\pm$ 0.21  & 25.60 $\pm$ 0.15  \nl
  2449857.013 & I & 25.06 $\pm$ 0.39  & 24.74 $\pm$ 0.12  & 23.77 $\pm$ 0.06  & 24.46 $\pm$ 0.16  & 24.70 $\pm$ 0.14  & 24.82 $\pm$ 0.17  \nl
  2449857.069 & I & 25.24 $\pm$ 0.23  & 25.00 $\pm$ 0.18  & 23.76 $\pm$ 0.08  & 24.29 $\pm$ 0.11  & 24.52 $\pm$ 0.34  & 24.56 $\pm$ 0.14  \nl
  2449857.082 & V & 25.89 $\pm$ 0.41  & 26.77 $\pm$ 0.95  & 24.64 $\pm$ 0.18  & 25.28 $\pm$ 0.34  & 26.62 $\pm$ 1.05  & 25.60 $\pm$ 0.24  \nl
  2449857.133 & I & 24.81 $\pm$ 0.30  & 24.78 $\pm$ 0.49  & 23.79 $\pm$ 0.16  & 24.50 $\pm$ 0.31  & 26.53 $\pm$ 0.78  & 25.05 $\pm$ 0.28  \nl
  2449862.174 & V & 25.46 $\pm$ 0.11  & 25.88 $\pm$ 0.13  & 24.88 $\pm$ 0.05  & 24.76 $\pm$ 0.14  & 25.17 $\pm$ 0.12  & 25.97 $\pm$ 0.13  \nl
  2449862.233 & V & 25.34 $\pm$ 0.11  & 25.98 $\pm$ 0.18  & 24.79 $\pm$ 0.08  & 24.73 $\pm$ 0.06  & 25.07 $\pm$ 0.09  & 25.94 $\pm$ 0.16  \nl
  2449868.206 & V & 25.40 $\pm$ 0.09  & 24.83 $\pm$ 0.06  & 25.18 $\pm$ 0.09  & 24.42 $\pm$ 0.04  & 25.70 $\pm$ 0.07  & 23.55 $\pm$ 0.24  \nl
  2449868.264 & V & 25.40 $\pm$ 0.22  & 24.74 $\pm$ 0.08  & 25.11 $\pm$ 0.11  & 24.46 $\pm$ 0.05  & 25.55 $\pm$ 0.12  & 26.31 $\pm$ 0.24  \nl
  2449874.974 & V & 26.02 $\pm$ 0.13  & 25.18 $\pm$ 0.09  & 25.45 $\pm$ 0.08  & 24.67 $\pm$ 0.08  & 26.22 $\pm$ 0.13  & 26.31 $\pm$ 0.17  \nl
  2449875.025 & V & 25.83 $\pm$ 0.17  & 25.11 $\pm$ 0.14  & 25.56 $\pm$ 0.11  & 24.65 $\pm$ 0.05  & 26.23 $\pm$ 0.23  & 25.79 $\pm$ 0.15  \nl
  2449883.417 & V & 26.40 $\pm$ 0.22  & 25.75 $\pm$ 0.13  & 24.48 $\pm$ 0.06  & 24.94 $\pm$ 0.10  & 25.14 $\pm$ 0.49  & 25.49 $\pm$ 0.13  \nl
  2449883.468 & V & 26.27 $\pm$ 0.16  & 25.36 $\pm$ 0.20  & 24.58 $\pm$ 0.08  & 25.05 $\pm$ 0.15  & 25.58 $\pm$ 0.13  & 25.59 $\pm$ 0.14  \nl
  2449883.482 & I & 25.17 $\pm$ 0.19  & 24.56 $\pm$ 0.12  & 23.71 $\pm$ 0.06  & 23.95 $\pm$ 0.11  & 24.55 $\pm$ 0.10  & 24.42 $\pm$ 0.13  \nl
  2449883.538 & I & 24.99 $\pm$ 0.19  & 24.43 $\pm$ 0.15  & 23.69 $\pm$ 0.06  & 22.87 $\pm$ 0.42  & 24.66 $\pm$ 0.09  & 24.81 $\pm$ 0.14  \nl
  2450203.095 & V & 26.22 $\pm$ 0.20  & 26.04 $\pm$ 0.17  & 24.40 $\pm$ 0.05  & 25.30 $\pm$ 0.13  & 25.35 $\pm$ 0.09  & 25.93 $\pm$ 0.20  \nl
  2450203.109 & V & 25.98 $\pm$ 0.15  & 26.12 $\pm$ 0.19  & 24.40 $\pm$ 0.08  & 25.06 $\pm$ 0.15  & 25.24 $\pm$ 0.08  & 26.44 $\pm$ 0.25  \nl
              &   &        C13        &        C14        &        C15        &        C16        &        C17        \nl
  2449819.813 & V & 25.35 $\pm$ 0.39  & 26.14 $\pm$ 0.33  & 26.67 $\pm$ 0.29  & 25.19 $\pm$ 0.07  & 25.47 $\pm$ 0.12  & 25.82 $\pm$ 0.11  \nl
  2449819.867 & V & 25.62 $\pm$ 0.17  & 26.15 $\pm$ 0.24  & 26.79 $\pm$ 0.38  & 25.33 $\pm$ 0.10  & 25.68 $\pm$ 0.18  & 25.92 $\pm$ 0.18  \nl
  2449828.528 & V & 25.71 $\pm$ 0.14  & 26.35 $\pm$ 0.20  & 25.95 $\pm$ 0.16  & 25.73 $\pm$ 0.11  & 26.04 $\pm$ 0.20  & 25.96 $\pm$ 0.10  \nl
  2449828.579 & V & 25.84 $\pm$ 0.17  & 23.33 $\pm$ 0.56  & 25.98 $\pm$ 0.16  & 25.83 $\pm$ 0.17  & 25.88 $\pm$ 0.11  & 25.04 $\pm$ 0.89  \nl
  2449828.593 & I & 24.69 $\pm$ 0.18  & 25.64 $\pm$ 0.27  & 25.05 $\pm$ 0.20  & 24.57 $\pm$ 0.19  & 24.67 $\pm$ 0.12  & 24.77 $\pm$ 0.69  \nl
  2449828.649 & I & 24.71 $\pm$ 0.11  & 25.57 $\pm$ 0.21  & 25.12 $\pm$ 0.14  & 24.38 $\pm$ 0.15  & 24.70 $\pm$ 0.08  & 24.95 $\pm$ 0.10  \nl
  2449839.777 & V & 25.18 $\pm$ 0.10  & 25.82 $\pm$ 0.14  & 26.42 $\pm$ 0.15  & 26.08 $\pm$ 0.09  & 24.89 $\pm$ 0.07  & 25.27 $\pm$ 0.10  \nl
  2449839.836 & V & 25.24 $\pm$ 0.11  & 25.55 $\pm$ 0.11  & 26.49 $\pm$ 0.22  & 26.20 $\pm$ 0.18  & 24.93 $\pm$ 0.09  & 25.38 $\pm$ 0.22  \nl
  2449839.850 & I & 24.25 $\pm$ 0.09  & 24.91 $\pm$ 0.12  & 25.17 $\pm$ 0.18  & 24.93 $\pm$ 0.17  & 23.99 $\pm$ 0.10  & 24.34 $\pm$ 0.12  \nl
  2449839.907 & I & 24.41 $\pm$ 0.09  & 24.88 $\pm$ 0.12  & 25.44 $\pm$ 0.25  & 24.70 $\pm$ 0.12  & 24.01 $\pm$ 0.08  & 24.21 $\pm$ 0.05  \nl
  2449842.722 & V & 25.14 $\pm$ 0.08  & 26.00 $\pm$ 0.16  & 25.88 $\pm$ 0.14  & 26.14 $\pm$ 0.13  & 25.12 $\pm$ 0.08  & 25.76 $\pm$ 0.08  \nl
  2449842.785 & V & 25.02 $\pm$ 0.09  & 25.88 $\pm$ 0.13  & 25.58 $\pm$ 0.10  & 26.22 $\pm$ 0.25  & 25.12 $\pm$ 0.09  & 25.60 $\pm$ 0.14  \nl
  2449845.269 & V & 25.21 $\pm$ 0.10  & 26.28 $\pm$ 0.15  & 26.30 $\pm$ 0.14  & 25.95 $\pm$ 0.16  & 25.21 $\pm$ 0.08  & 25.80 $\pm$ 0.12  \nl
  2449845.288 & V & 25.07 $\pm$ 0.11  & 26.41 $\pm$ 0.40  & 26.20 $\pm$ 0.19  & 26.34 $\pm$ 0.24  & 25.14 $\pm$ 0.11  & 25.56 $\pm$ 0.07  \nl
  2449848.756 & V & 25.40 $\pm$ 0.11  & 26.57 $\pm$ 0.16  & 26.60 $\pm$ 0.20  & 26.09 $\pm$ 0.16  & 25.40 $\pm$ 0.08  & 26.00 $\pm$ 0.14  \nl
  2449848.819 & V & 25.30 $\pm$ 0.05  & 26.99 $\pm$ 0.38  & 27.21 $\pm$ 0.40  & 26.01 $\pm$ 0.22  & 25.43 $\pm$ 0.13  & 25.42 $\pm$ 0.35  \nl
  2449852.993 & V & 25.51 $\pm$ 0.09  & 26.66 $\pm$ 0.29  & 26.63 $\pm$ 0.22  & 25.52 $\pm$ 0.13  & 25.71 $\pm$ 0.09  & 26.02 $\pm$ 0.17  \nl
  2449853.044 & V & 25.60 $\pm$ 0.15  & 26.36 $\pm$ 0.21  & 26.83 $\pm$ 0.33  & 25.47 $\pm$ 0.10  & 24.64 $\pm$ 0.40  & 25.86 $\pm$ 0.10  \nl
  2449856.946 & V & 25.58 $\pm$ 0.15  & 25.53 $\pm$ 0.11  & 26.06 $\pm$ 0.16  & 25.41 $\pm$ 0.09  & 25.80 $\pm$ 0.09  & 25.93 $\pm$ 0.08  \nl
  2449856.999 & V & 25.61 $\pm$ 0.14  & 25.70 $\pm$ 0.13  & 25.84 $\pm$ 0.14  & 25.38 $\pm$ 0.09  & 25.69 $\pm$ 0.15  & 26.29 $\pm$ 0.19  \nl
  2449857.013 & I & 24.29 $\pm$ 0.11  & 24.84 $\pm$ 0.16  & 25.00 $\pm$ 0.21  & 24.39 $\pm$ 0.14  & 24.48 $\pm$ 0.09  & 24.75 $\pm$ 0.13  \nl
  2449857.069 & I & 24.38 $\pm$ 0.11  & 24.94 $\pm$ 0.14  & 25.41 $\pm$ 0.16  & 24.51 $\pm$ 0.08  & 24.56 $\pm$ 0.09  & 24.72 $\pm$ 0.18  \nl
  2449857.082 & V & 25.65 $\pm$ 0.33  & 25.99 $\pm$ 0.46  & 25.85 $\pm$ 0.41  & 25.21 $\pm$ 0.13  & 21.94 $\pm$ 0.36  & 25.81 $\pm$ 0.41  \nl
  2449857.133 & I & 24.67 $\pm$ 0.42  & 25.32 $\pm$ 0.57  & 24.92 $\pm$ 0.35  & 25.23 $\pm$ 0.74  & 24.91 $\pm$ 0.32  & 25.27 $\pm$ 0.45  \nl
  2449862.174 & V & 25.75 $\pm$ 0.14  & 26.42 $\pm$ 0.18  & 26.66 $\pm$ 0.32  & 25.51 $\pm$ 0.09  & 25.72 $\pm$ 0.09  & 24.84 $\pm$ 0.06  \nl
  2449862.233 & V & 25.78 $\pm$ 0.17  & 26.28 $\pm$ 0.28  & 26.77 $\pm$ 0.34  & 25.56 $\pm$ 0.12  & 25.75 $\pm$ 0.70  & 24.80 $\pm$ 0.07  \nl
  2449868.206 & V & 25.85 $\pm$ 0.15  & 26.67 $\pm$ 0.56  & 26.59 $\pm$ 0.15  & 25.86 $\pm$ 0.14  & 24.64 $\pm$ 0.07  & 25.30 $\pm$ 0.08  \nl
  2449868.264 & V & 25.85 $\pm$ 0.16  & 26.77 $\pm$ 0.30  & 26.51 $\pm$ 0.24  & 25.56 $\pm$ 0.29  & 24.70 $\pm$ 0.07  & 25.29 $\pm$ 0.09  \nl
  2449874.974 & V & 25.47 $\pm$ 0.13  & 25.71 $\pm$ 0.12  & 26.42 $\pm$ 0.25  & 26.14 $\pm$ 0.14  & 25.11 $\pm$ 0.08  & 25.60 $\pm$ 0.12  \nl
  2449875.025 & V & 25.47 $\pm$ 0.15  & 25.75 $\pm$ 0.13  & 26.25 $\pm$ 0.20  & 26.14 $\pm$ 0.14  & 25.16 $\pm$ 0.08  & 25.69 $\pm$ 0.10  \nl
  2449883.417 & V & 25.05 $\pm$ 0.57  & 25.97 $\pm$ 0.47  & 26.10 $\pm$ 0.14  & 26.12 $\pm$ 0.17  & 25.59 $\pm$ 0.11  & 26.02 $\pm$ 0.13  \nl
  2449883.468 & V & 25.22 $\pm$ 0.09  & 26.92 $\pm$ 0.27  & 26.20 $\pm$ 0.22  & 26.06 $\pm$ 0.22  & 25.59 $\pm$ 0.13  & 25.73 $\pm$ 0.12  \nl
  2449883.482 & I & 24.07 $\pm$ 0.08  & 25.26 $\pm$ 0.35  & 25.64 $\pm$ 0.33  & 25.19 $\pm$ 0.22  & 24.36 $\pm$ 0.09  & 24.65 $\pm$ 0.13  \nl
  2449883.538 & I & 24.18 $\pm$ 0.09  & 25.42 $\pm$ 0.20  & 25.19 $\pm$ 0.21  & 25.31 $\pm$ 0.18  & 23.07 $\pm$ 0.31  & 24.81 $\pm$ 0.10  \nl
  2450203.095 & V & 25.79 $\pm$ 0.11  & 26.62 $\pm$ 0.38  & 26.53 $\pm$ 0.22  & 26.23 $\pm$ 0.16  & 25.88 $\pm$ 0.14  & 26.00 $\pm$ 0.14  \nl
  2450203.109 & V & 25.69 $\pm$ 0.12  & 27.22 $\pm$ 0.37  & 26.96 $\pm$ 0.28  & 26.27 $\pm$ 0.25  & 25.75 $\pm$ 0.12  & 26.15 $\pm$ 0.24  \nl
              &   &        C19        &        C20        &                   &                   &                   &                   \nl
  2449819.813 & V & 25.35 $\pm$ 0.08  & 26.47 $\pm$ 0.18  &                   &                   &                   &                   \nl
  2449819.867 & V & 25.32 $\pm$ 0.10  & 26.93 $\pm$ 0.39  &                   &                   &                   &                   \nl
  2449828.528 & V & 25.57 $\pm$ 0.08  & 26.08 $\pm$ 0.11  &                   &                   &                   &                   \nl
  2449828.579 & V & 25.55 $\pm$ 0.16  & 26.13 $\pm$ 0.19  &                   &                   &                   &                   \nl
  2449828.593 & I & 22.99 $\pm$ 0.34  & 25.06 $\pm$ 0.16  &                   &                   &                   &                   \nl
  2449828.649 & I & 24.30 $\pm$ 0.06  & 25.48 $\pm$ 0.19  &                   &                   &                   &                   \nl
  2449839.777 & V & 25.76 $\pm$ 0.14  & 25.78 $\pm$ 0.15  &                   &                   &                   &                   \nl
  2449839.836 & V & 25.58 $\pm$ 0.12  & 25.84 $\pm$ 0.14  &                   &                   &                   &                   \nl
  2449839.850 & I & 24.32 $\pm$ 0.13  & 25.27 $\pm$ 0.21  &                   &                   &                   &                   \nl
  2449839.907 & I & 24.29 $\pm$ 0.09  & 25.47 $\pm$ 0.21  &                   &                   &                   &                   \nl
  2449842.722 & V & 25.90 $\pm$ 0.14  & 26.35 $\pm$ 0.17  &                   &                   &                   &                   \nl
  2449842.785 & V & 25.76 $\pm$ 0.15  & 26.12 $\pm$ 0.20  &                   &                   &                   &                   \nl
  2449845.269 & V & 25.77 $\pm$ 0.16  & 26.44 $\pm$ 0.13  &                   &                   &                   &                   \nl
  2449845.288 & V & 25.77 $\pm$ 0.18  & 26.50 $\pm$ 0.22  &                   &                   &                   &                   \nl
  2449848.756 & V & 25.67 $\pm$ 0.14  & 26.58 $\pm$ 0.19  &                   &                   &                   &                   \nl
  2449848.819 & V & 25.43 $\pm$ 0.10  & 26.84 $\pm$ 0.28  &                   &                   &                   &                   \nl
  2449852.993 & V & 25.43 $\pm$ 0.13  & 25.50 $\pm$ 0.09  &                   &                   &                   &                   \nl
  2449853.044 & V & 25.53 $\pm$ 0.11  & 25.59 $\pm$ 0.11  &                   &                   &                   &                   \nl
  2449856.946 & V & 25.27 $\pm$ 0.07  & 26.16 $\pm$ 0.11  &                   &                   &                   &                   \nl
  2449856.999 & V & 24.09 $\pm$ 0.45  & 25.91 $\pm$ 0.15  &                   &                   &                   &                   \nl
  2449857.013 & I & 24.18 $\pm$ 0.07  & 25.25 $\pm$ 0.19  &                   &                   &                   &                   \nl
  2449857.069 & I & 24.15 $\pm$ 0.07  & 25.22 $\pm$ 0.14  &                   &                   &                   &                   \nl
  2449857.082 & V & 25.92 $\pm$ 0.41  & 26.60 $\pm$ 0.53  &                   &                   &                   &                   \nl
  2449857.133 & I & 24.32 $\pm$ 0.17  & 26.06 $\pm$ 1.41  &                   &                   &                   &                   \nl
  2449862.174 & V & 25.26 $\pm$ 0.08  & 26.65 $\pm$ 0.19  &                   &                   &                   &                   \nl
  2449862.233 & V & 25.16 $\pm$ 0.08  & 26.58 $\pm$ 0.24  &                   &                   &                   &                   \nl
  2449868.206 & V & 25.30 $\pm$ 0.10  & 25.85 $\pm$ 0.11  &                   &                   &                   &                   \nl
  2449868.264 & V & 25.45 $\pm$ 0.14  & 26.02 $\pm$ 0.13  &                   &                   &                   &                   \nl
  2449874.974 & V & 25.53 $\pm$ 0.11  & 26.40 $\pm$ 0.23  &                   &                   &                   &                   \nl
  2449875.025 & V & 25.36 $\pm$ 0.11  & 26.81 $\pm$ 0.47  &                   &                   &                   &                   \nl
  2449883.417 & V & 25.66 $\pm$ 0.11  & 26.03 $\pm$ 0.13  &                   &                   &                   &                   \nl
  2449883.468 & V & 25.59 $\pm$ 0.16  & 26.29 $\pm$ 0.16  &                   &                   &                   &                   \nl
  2449883.482 & I & 24.32 $\pm$ 0.10  & 25.09 $\pm$ 0.19  &                   &                   &                   &                   \nl
  2449883.538 & I & 24.42 $\pm$ 0.08  & 25.21 $\pm$ 0.22  &                   &                   &                   &                   \nl
  2450203.095 & V & 25.30 $\pm$ 0.10  & 23.73 $\pm$ 0.52  &                   &                   &                   &                   \nl
  2450203.109 & V & 25.36 $\pm$ 0.10  & 26.04 $\pm$ 0.14  &                   &                   &                   &                   \nl
\enddata
\end{deluxetable}

\clearpage

\begin{deluxetable}{clcl}
\footnotesize
\tablecaption{ALLFRAME Error Budget\tablenotemark{ }
\label{tbl:error}}
\tablewidth{0pt}
\tablehead{
\colhead{ } & \colhead{Source of Uncertainty} & \colhead{Error (mag)} &
\colhead{\quad Notes}
}
\startdata
\multicolumn{2}{c}{\bf CEPHEID PL CALIBRATION} & & \nl
(a)   & LMC True Modulus            & $\pm$0.10  & (1) \nl
(b)   & V PL Zero Point             & $\pm$0.05  & (2),(3) \nl
(c)   & I PL Zero Point             & $\pm$0.03  & (2),(4) \nl
(S1)  & PL Systematic Uncertainty   & $\pm$0.12  & (a),(b),(c) combined in quadrature \nl
\tablevspace{6pt}
\multicolumn{2}{c}{\bf NGC~4725 MODULUS} & & \nl
(d)   & HST V-Band Zero Point       & $\pm$0.05  & (5) \nl
(e)   & HST I-Band Zero Point       & $\pm$0.05  & (5) \nl
(R1)  & Cepheid True Modulus        & $\pm$0.15  & (6) \nl
(R2)  & Dereddened PL Fit            & $\pm$0.06  & (7) \nl
(S2)  & Metallicity Uncertainty     & $+0.12\pm 0.21$ & See text for details \nl
\tablevspace{6pt}
\multicolumn{2}{c}{\bf TOTAL UNCERTAINTY} & & \nl
(R)   & Random Errors               & $\pm$0.16  & (R1),(R2) combined in quadrature \nl
(S)   & Systematic Errors           & $\pm$0.17  & (S1),(S2) combined in quadrature \nl
\enddata
\tablenotetext{ }
{
\hspace{-7.0mm} (1) Adopted from Madore \& Freedman (1991).
(2) Derived from the observed scatter in the Madore \& Freedman (1991) PL
relation, with 32 contributing Cepheids.
(3) V-band $1\sigma$ scatter: $\pm 0.27$ mag.
(4) I-band $1\sigma$ scatter: $\pm 0.18$ mag.
(5) Contributing uncertainties from aperture corrections, the Holtzmann \etal
(1995) zero points, and the long versus short uncertainty, combined in
quadrature.  Adopted aperture correction contribution is the worst-case formal
uncertainty ($\pm 0.04$ mag) for the NGC~4725 aperture corrections.  Adopted
Holtzmann \etal zero point uncertainty is $\pm 0.02$ mag.  Adopted long versus
short exposure correction uncertainty is $\pm 0.02$ mag.
(6) Assuming that photometric errors (d,e) are uncorrelated between filters,
and noting that that V and I magnitudes are multiplied by +1.45 and -2.45,
respectively, when correcting for reddening, results in a derived error on the
true modulus of $[(1.45)^2(0.05)^2+(-2.45)^2(0.05)^2]^{1/2}=0.15$ mag.
(7) Uncertainties for the mean true modulus result from the finite width of the
instability strip and the random star-to-star photometric errors,
reduced by the population size of
contributing Cepheids for NGC~4725 (20 variables).  
}
\end{deluxetable}

\clearpage

\begin{deluxetable}{lll}
\footnotesize
\tablecaption{Published Distances to NGC~4725 and the Coma Cloud\tablenotemark{a}
\label{tbl:distances}}
\tablewidth{0pt}
\tablehead{
\colhead{Method} & \colhead{Distance (Mpc)} & \colhead{Reference}
}
\startdata
\multicolumn{3}{c}{\it NGC~4725}\nl
TF (B-band)		& $\;\; 9.9 \pm 1.0$  & Bottinelli \etal (1985)	\nl
Mass Model   		& $12.4$	  & Tully (1988)		\nl
TF (H-band)		& $16.1$	  & Tully \etal (1992)		\nl
Mass Model 		& $20$	  	  & Tully \etal (1992)		\nl
TF (BRIH-band)		& $12.6 \pm 2.1$  & Tully (1997)		\nl
SBF                 	& $13.1 \pm 2.2$  & Tonry (1998)		\nl
Cepheids		& $12.6 \pm 1.0$  & This paper (ALLFRAME)     	\nl\nl
\multicolumn{3}{c}{\it NGC~4414}\nl
Cepheids		& $19.1 \pm 1.6$  & Turner \etal (1998)       	\nl\nl
\multicolumn{3}{c}{\it NGC~4278}\nl
GCLF\tablenotemark{b}   & $13.2 \pm 0.9$  & Forbes  (1996)		\nl
PNLF                 	& $10.2 \pm 1.0$  & Jacoby \etal (1996)		\nl\nl
\multicolumn{3}{c}{\it NGC~4494}\nl
Mass Model 		& $11.7$ 	  & Tully \& Shaya (1984)       \nl
SBF                 	& $15.0 \pm 2.3$  & Simard \& Pritchet (1994)	\nl
GCLF\tablenotemark{b}	& $14.5 \pm 2.9$  & Fleming \etal (1995)	\nl
GCLF\tablenotemark{b}   & $12.6 \pm 0.9$  & Forbes  (1996)		\nl
PNLF                 	& $12.8 \pm 0.9$  & Jacoby \etal (1996)		\nl\nl
\multicolumn{3}{c}{\it NGC~4565}\nl
Mass Model 		& $11.0$ 	  & Tully \& Shaya (1984)       \nl
SBF                 	& $10.4 \pm 0.4$  & Simard \& Pritchet (1994)	\nl
GCLF\tablenotemark{b}	& $10.0 \pm 1.5$  & Fleming \etal (1995)	\nl
PNLF                 	& $10.5 \pm 1.0$  & Jacoby \etal (1996)		\nl\nl
\multicolumn{3}{c}{\it Average of NGC~4150,4251,4283}\nl
SBF                 	& $15.5 \pm 0.6$  & Tonry \etal (1997)		\nl\nl
\multicolumn{3}{c}{\it Average of NGC~4494,4565,4725}\nl
SBF\tablenotemark{c}   	& $15.9 \pm 0.6$  & Tonry \etal (1997)		\nl
\enddata
\tablenotetext{a}{\scriptsize 
For convenience, the presented Coma Cloud inventory
follows that of Tully (1988) - \ie his Groups 14-1 (Coma I) and 14-2 (Coma II), 
respectively, from his Table II.  The complexity of this region makes any
unequivocal claim of a true physical association between \it all \rm (or,
arguably, any) of the listed Coma Cloud ``members'' highly suspect.}
\tablenotetext{b}{\scriptsize
Fleming et~al.'s (1995) results are 
based upon ground-based CFHT data,
whereas Forbes (1996) used $HST$.  
The latter reference revisits Fleming et~al.'s
conclusions, in light of the $HST$ results.}
\tablenotetext{c}{\scriptsize
Tonry \etal (1997) compute a Coma II Group distance based
upon the (approximate) mean of the SBF distances to NGC~4494, 4565, and 4725
(Tonry 1998).  Tully's (1988) inventory would place NGC~4494
and 4565 in the Coma I Group, with only NGC~4725 \it strictly \rm a Coma II
Group member.}
\end{deluxetable}

\clearpage



















\epsscale{1.0}
\plotone{f5.eps}

\clearpage

\epsscale{1.0}
\plotone{f6.eps}

\clearpage

\epsscale{1.0}
\plotone{f7.eps}

\clearpage

\epsscale{1.0}
\plotone{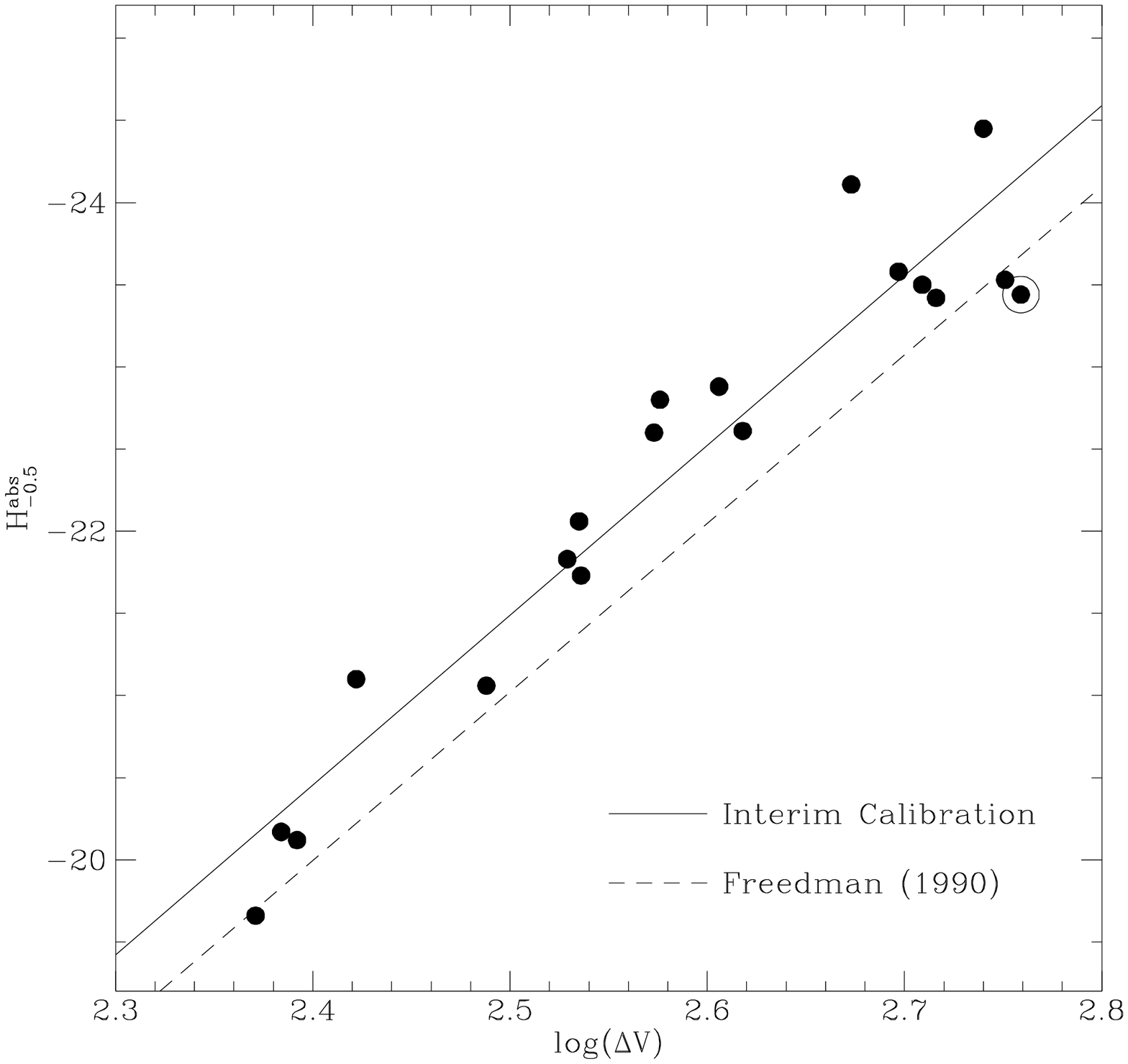}

\end{document}